\documentclass[aip,jcp,amsmath,amssymb,floatfix,citeautoscript,reprint]{revtex4-1}
\usepackage{graphicx}
\usepackage{dcolumn}
\usepackage{bm}
\usepackage{braket}
\usepackage{mathtools}
\usepackage{xcolor}
\usepackage{mathrsfs}

\newcommand{\cm}{\mathrm{cm}^{-1}}

\newif\ifhighlightnumbers

\newcommand{\numericalresult}[1]{\ifhighlightnumbers \textcolor{red}{#1}\else #1\fi}

\begin{document}
\title{Two-dimensional electronic spectroscopy in the condensed phase using equivariant transformer accelerated molecular dynamics simulations}

\author{Joseph Kelly}
\affiliation{Department of Chemistry, Stanford University, Stanford, California, 94305, USA}

\author{Frank Hu}
\affiliation{Department of Chemistry, Stanford University, Stanford, California, 94305, USA}

\author{Arianna Damiani}
\affiliation{Department of Chemistry, Stanford University, Stanford, California, 94305, USA}

\author{Michael S. Chen}
\affiliation{Simons Center for Computational Physical Chemistry, Department of Chemistry, New York University, New York, New York 10003, United States}

\author{Andrew Snider}
\affiliation{Department of Chemistry and Biochemistry, University of California Merced, Merced, California 95343, USA}

\author{Minjung Son}
\affiliation{Department of Chemistry, Boston University, Boston, Massachusetts 02215, USA}

\author{Angela Lee}
\affiliation{Department of Chemistry, Massachusetts Institute of Technology, Cambridge, Massachusetts 02139, USA}

\author{Prachi Gupta}
\affiliation{Department of Chemistry and Biochemistry, University of California Merced, Merced, California 95343, USA}

\author{Andr\'es Montoya-Castillo}
\affiliation{Department of Chemistry, University of Colorado, Boulder, Boulder, Colorado, 80309, USA}

\author{Tim J. Zuehlsdorff}
 \affiliation{Department of Chemistry, Oregon State University, Corvallis, Oregon 97331, USA}

\author{Gabriela S. Schlau-Cohen}
\email{gssc@mit.edu}
\affiliation{Department of Chemistry, Massachusetts Institute of Technology, Cambridge, Massachusetts 02139, USA}

\author{Christine M. Isborn}
\email{cisborn@ucmerced.edu}
\affiliation{Department of Chemistry and Biochemistry, University of California Merced, Merced, California 95343, USA}

\author{Thomas E. Markland}
\email{tmarkland@stanford.edu}
\affiliation{Department of Chemistry, Stanford University, Stanford, California, 94305, USA}

\date{\today}

\begin{abstract}
Two-dimensional electronic spectroscopy (2DES) provides rich information about how the electronic states of molecules, proteins, and solid-state materials interact with each other and their surrounding environment. Atomistic molecular dynamics simulations offer an appealing route to uncover how nuclear motions mediate electronic energy relaxation and their manifestation in electronic spectroscopies, but are computationally expensive. Here we show that, by using an equivariant transformer-based machine learning architecture trained with only ~2500 ground state and ~100 excited state electronic structure calculations, one can construct accurate machine-learned potential energy surfaces for both the ground-state electronic surface and excited-state energy gap. We demonstrate the utility of this approach for simulating the dynamics of Nile blue in ethanol, where we experimentally validate and decompose the simulated 2DES to establish the nuclear motions of the chromophore and the solvent that couple to the excited state, connecting the spectroscopic signals to their molecular origin.
\end{abstract}

\maketitle
\normalsize

Extracting information about the interplay between electronic and nuclear dynamics in complex condensed phase materials, ranging from light-harvesting complexes to quantum dots, is often greatly enhanced by going beyond linear electronic spectroscopy. Two-dimensional electronic spectroscopy\cite{Hybl1998Two-dimensionalSpectroscopy, Hybl2001Two-dimensionalSpectroscopy} (2DES) is a nonlinear spectroscopic technique that provides access to features that encode population dynamics within and between states as well as the couplings of these states distributed across a range of excitation and emission frequencies. Characterizing these features and attributing them to specific molecular motions, particularly in condensed phase environments where these can be highly influenced by the structure and dynamics of the environment (e.g. solvent), is challenging because of the complexity of these systems. Understanding these observations has been deeply enriched by leveraging atomistic simulation. It is therefore important to develop efficient simulation methods that can accurately predict 2DES signals capturing phenomena such as anharmonic and vibronic effects, thermal broadening, and electronic interactions with the environment. Capturing these effects requires generating atomistic molecular dynamics (MD) on timescales of 50 to 100 picoseconds, along with the corresponding electronic excitation energies, to sample the nuclear fluctuations of both the chromophore and its environment. Assuming an {\it ab initio} treatment of both the dynamics and the environment, achieving this requires hundreds of thousands of ground and excited state electronic structure calculations for systems that may exceed 500 atoms owing to the desire to include potential charge transfer and polarization between the chromophore and its condensed phase environment. Such a large number of calculations is typically intractable at anything but the lowest levels of electronic structure theory, such as semiempirical methods.

Recent work has suggested opportunities to use machine learning (ML) approaches to provide a tractable route to efficiently compute and understand electronic properties ranging from electronic excitation energies in the gas phase to condensed phase linear and 2D electronic spectra. For gas-phase systems these ML approaches have been used to predict orbital energy gaps\cite{Fuchs2020SE3-Transformers:Networks,Luo2023OneData,Liao2023Equiformer:Graphs,Cignoni2024ElectronicLearning} and excitation energies\cite{Ramakrishnan2015ElectronicSpace,Pronobis2018CapturingLearning,Ghosh2019DeepSpectra,Mazouin2022SelectedData-efficiency,Entwistle2023ElectronicCarlo,Baker2024InvariantMolecules} for databases of molecules at their optimized geometries. To obtain gas phase linear spectra that incorporate the shape of the electronic absorption peaks arising from molecular geometries beyond their optimized structure, recent work has trained ML models to predict electronic energy gaps, i.e. the difference in energy between the electronic states, at multiple geometries and combined these with the ensemble approach\cite{Mukamel1995PrinciplesSpectroscopy,Crespo-Otero2012Spectrum2-phenylfuran,Isborn2012ElectronicProtein,Ge2015AccurateSolution,Marenich2015ElectronicSampling,Milanese2017ConvergenceSolvent,Zuehlsdorff2017PredictingRed} to compute linear spectra \cite{Ye2019AN-methylacetamide,Xue2020MachineSections,Lu2020DeepSemiconductors,Westermayr2020DeepSpace,Zou2023ASpectra,Vinod2023MultifidelityEnergies,Petrusevich2023Cost-EffectiveBroadening}. ML potential energy surfaces have also been trained to the quality required to perform nonadiabatic dynamics simulations to study electronic relaxation through conical intersections in gas phase systems \cite{Chen2018DeepDynamics,Westermayr2019MachineSimulations,Axelrod2022ExcitedPotential,Shakiba2024Machine-LearnedDynamics}. Machine learned potentials also show considerable promise in condensed phase studies such as in the construction of model Hamiltonians for large networks of coupled chromophores \cite{Hase2016MachineDynamics,Farahvash2020MachineDynamics,Cignoni2023MachineComplexes,Betti2024InsightsSimulations}, calculating ensemble approach linear electronic absorption spectra by machine learning electronic energy gaps in implicit solvent \cite{Bononi2020BathochromicCalculations,Chen2022UV-VisibleLearning}, and using electrostatic machine learning embedding to accurately capture environmental and vibronic effects\cite{Zinovjev2025ImprovedPotentials}. 

Our recent work has sought to employ ML models to compute the excited state electronic energy gaps for chromophores in condensed phase environments enabling the combination of a dynamical treatment of linear and 2D electronic spectra via the second-order cumulant 
approach\cite{Mukamel1985FluorescenceEigenstates,Mukamel1995PrinciplesSpectroscopy,Zuehlsdorff2019OpticalApproaches}, allowing for the inclusion of vibronic and anharmonic effects \cite{Zuehlsdorff2019OpticalApproaches}, combined with an explicit treatment of the solvent\cite{Chen2020ExploitingEnvironments}. This work has allowed us to show that one can train an accurate ML model of the electronic energy gap using only $\sim$2000 energies \cite{Chen2020ExploitingEnvironments} or just $\sim$300 electronic energy gap gradients \cite{Chen2023ElucidatingStructure}. By exploiting transfer learning we have also shown that, starting from ML energy gap surfaces trained using computationally affordable excited-state electronic structure methods including time-dependent density-functional theory (TDDFT) or configuration interaction singles (CIS), one can adapt the model to the more accurate but more expensive equation-of-motion coupled cluster with singles and doubles (EOM-CCSD) level of theory using only 100s of energies\cite{Chen2023ElucidatingStructure}. This has allowed us to show that the explicit treatment of the hydrogen bonds (H-bonds) to the anionic green fluorescent protein chromophore in liquid water combined with a high-level wavefunction-based description of the excited state electronic structure is essential to capture the experimentally observed breadth of the linear electronic absorption spectrum and that this, in turn, leads to large changes in the 2DES signals \cite{Chen2023ElucidatingStructure}.

Here we show that using an Equivariant Transformer (EqT) architecture \cite{Vaswani2017AttentionNeed,Tholke2021EquivariantPotentials,Pelaez2024TorchMD-NetSimulations}, which naturally encodes permutational, rotational, and translational symmetries within the machine-learned model, one can capture both the ground state electronic potential energy and electronic energy gap surfaces to simulate atomistic MD, the excitation energy gaps, and the corresponding linear and 2D electronic absorption spectra of a solvated chromophore using the energies and gradients obtained from only \numericalresult{2546} ground state and \numericalresult{90} excited state electronic structure calculations (SI Sec.~IB-ID) compared to $\sim$100,000 of both without the use of ML. We demonstrate this approach by simulating the electronic spectroscopy of Nile blue in ethanol and comparing it to recent 2DES experiments\cite{Son2017UltrabroadbandDetection}. In doing so, we reveal how atomistic MD simulations can be used to unravel the rich information present in 2DES signals and highlight the additional understanding that one can extract about the role of the atomistic environment. The EqT approach offers significant improvements over our previous work as an EqT model trained with \numericalresult{less than 100} excited state calculations can achieve a root mean squared error (RMSE) of \numericalresult{less than 10 meV} compared to 1000's of training calculations achieving RMSEs of 50-80 meV using the atom-centered neural network architecture based on Chebyshev polynomial descriptors\cite{Artrith2017EfficientSpecies} that we previously employed\cite{Chen2020ExploitingEnvironments}. The EqT architecture thus offers an extremely data-efficient route to fully atomistic, condensed-phase predictions of linear and 2DES spectra. Avoiding the tour-de-force {\it ab initio} molecular dynamics and excited state electronic structure calculations previously required will enable tighter feedback loops between experiment and simulation (SI Sec.~IE). Upon establishing the efficiency of this approach, we then compare our simulations of the linear, 2DES, and pump-probe spectra to recent experiments of Nile blue in ethanol to unravel how solvation both drives electronic energy relaxation and enhances coupling to specific chromophore vibrational motions. 

\begin{figure}[h]
    \begin{center}
        \includegraphics[width=0.45\textwidth]{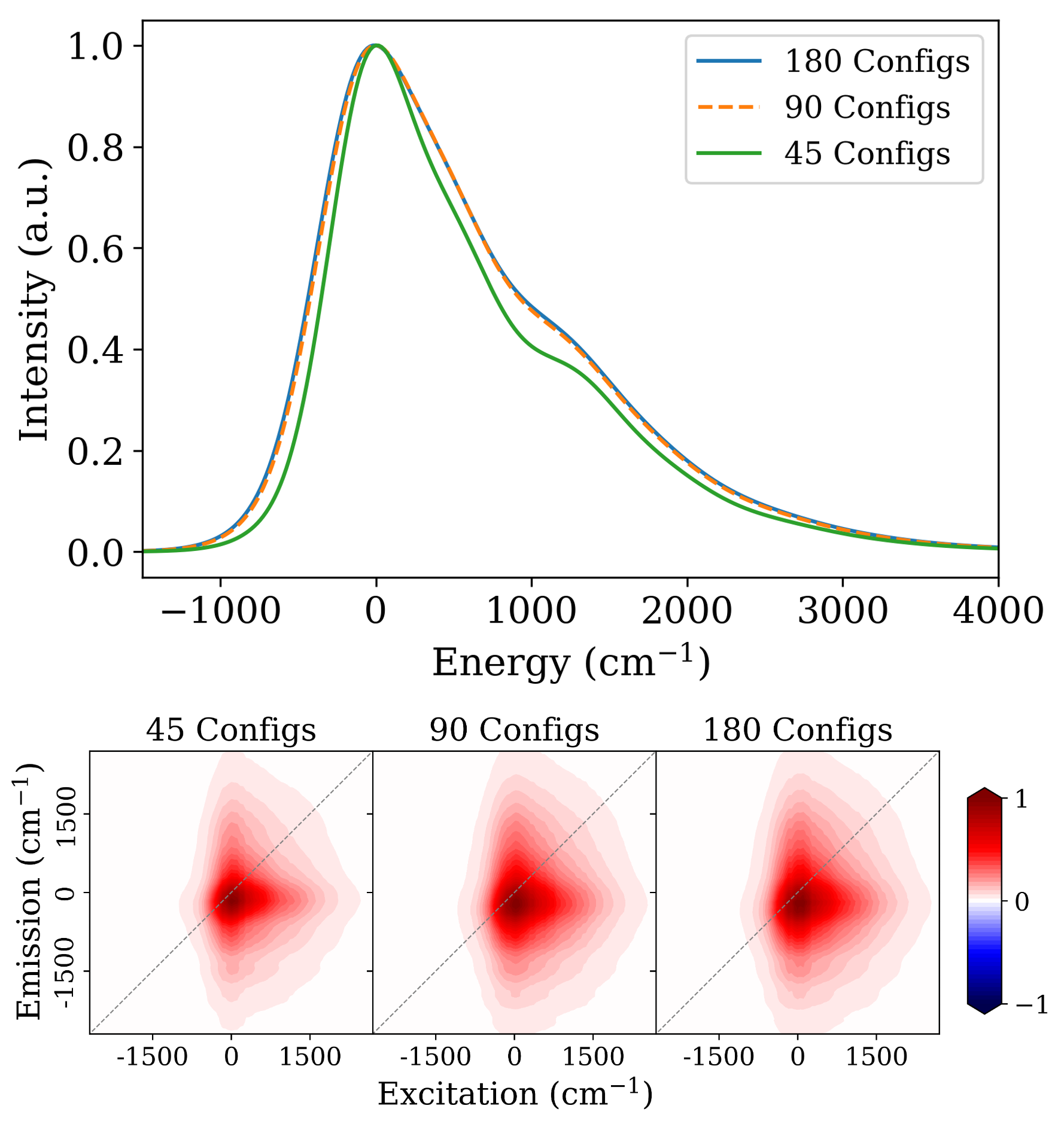}
    \end{center}
    \vspace{-5mm}
    \caption{Convergence of the spectra predicted by the EqT models of the electronic excitation energy gap. Top: Simulated linear electronic absorption spectrum of Nile blue in ethanol as a function of the number of configurations (configs) used to train the EqT electronic energy gap model. The intensity maxima of all the spectra are shifted to be at 0 cm$^{-1}$ to allow for easier comparison of convergence of the line shapes. The 90 and 180 configuration models produce graphically indistinguishable line shapes. Bottom: Simulated 2DES spectra with experimental pulse spectral profile applied (SI Sec.~II) at a time delay of $t_2=600~{\rm fs}$ as a function of the number of training configurations used for the EqT energy gap model. The zero frequency of the x and y-axes are defined in the same way as the linear spectra above. Each of the three models give very similar spectra with only a minimal increase in breadth noticeable from 45 to 90 configurations.} 
    \label{fig:energyMLconv}    
\end{figure}

To calculate linear and 2DES spectra, we use the second-order cumulant approach, \cite{Mukamel1985FluorescenceEigenstates,Mukamel1995PrinciplesSpectroscopy,Zuehlsdorff2019OpticalApproaches} which requires obtaining a one-time correlation function of electronic energy gaps from an MD trajectory. To train an ML surface on which to perform the ground state dynamics, we began with a 50 ps density-functional tight-binding (DFTB)\cite{Porezag1994ConstructionCarbon, Seifert1996CalculationsScheme} trajectory of Nile blue solvated by 175 ethanol molecules, selected configurations, calculated their energies and gradients with DFT using the revPBE exchange-correlation functional, trained an EqT model, and then iteratively added more configurations until the machine-learned potential was stable and achieving a gradient RMSE error on the test set of 44 meV/Å (SI Sec.~IB). To calculate the ground to excited state energy difference, we sampled configurations from the ground-state EqT dynamics, calculated CAM-B3LYP/6-31G* excitation energies and excited state gradients, treating the Nile blue and, on average, 118 ethanol molecules at the same level of theory using TeraChem v1.9 \cite{Seritan2021TeraChem:Dynamics} (SI Sec.~IC), and used these values to train a second EqT model. We used the final ground state EqT model to generate 1~ns of dynamics then used the excitation energy EqT model to predict the energy gaps for every configuration in that trajectory, yielding a trajectory of 500,000 energy gap predictions trained using fewer than 200 excited state electronic structure calculations (SI Sec.~ID).

Fig.~\ref{fig:energyMLconv} shows the convergence of the simulated linear absorption and 2DES spectra for Nile blue in ethanol as the number of configurations used to train the EqT electronic energy gap are increased. From this, one can see that for the linear absorption spectra (top), even when only \numericalresult{45} configurations are used, the shape of the spectrum is close to convergence while the results obtained from \numericalresult{90 and 180} training configurations are graphically indistinguishable. The RMSE on the test set falls from \numericalresult{30} meV to \numericalresult{24} meV to \numericalresult{12} meV as the number of training points is increased across this range, suggesting that errors below \numericalresult{$\sim$20 meV} in the energy gap manifest negligibly in the linear spectrum. This is not surprising given that the width of the absorption spectrum spans 0.6 eV, which is \numericalresult{50} times larger than the RMSE in the EqT model with \numericalresult{180} training configurations. The bottom panel of Fig.~\ref{fig:energyMLconv} shows the convergence of the 2DES as the number of configurations on which the electronic energy gap model was trained is increased. Similar to the linear absorption spectrum, rapid convergence is observed with the model trained on 45 configurations, giving a slightly narrower 2D spectrum and those with 90 and 180 configurations being indistinguishable. The EqT approach therefore allows one to use $\sim$100 excited state electronic structure calculations to build an electronic energy gap model that yields converged linear and 2D electronic spectra.

\begin{figure}[h]
    \begin{center}
        \includegraphics[width=0.45\textwidth]{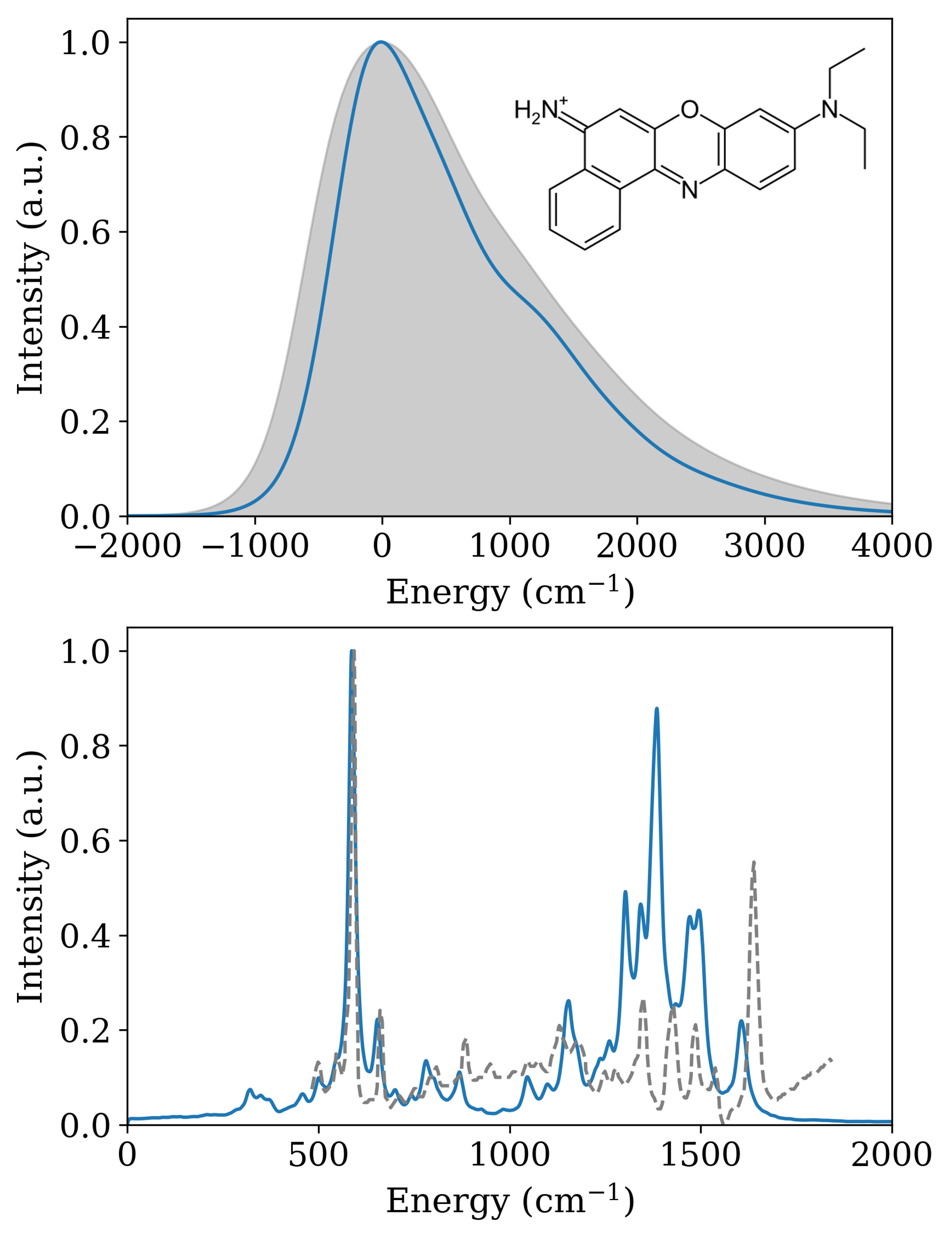}
    \end{center}
    \vspace{-5mm}
    \caption{Comparison of simulated linear spectrum with experiment and spectral density with SERS. Top: Experimentally obtained (gray, SI Sec.~IA) and simulated (blue) linear absorption spectra of the Nile blue chromophore in ethanol. Shown in the inset is the chemical structure of the Nile blue cation. Experiment and simulation are separately shifted (by 1.97 and 2.29 eV) so their peak maxima match at 0 $\cm$ to allow comparison of the spectral line shapes. Bottom: Simulated spectral density ($J(\omega)$, blue) indicates which nuclear motions couple to the electronic excitation and experimentally obtained SERS signal\cite{Auguie2012TinyAbsorptions} (gray) as a comparison. The SERS spectrum is shifted vertically so all signal is non-negative then normalized to a maximum value of one.} 
    \label{fig:linspec}
\end{figure}

Given the converged spectra obtained from the EqT model, we now compare these spectra with recent experiments. The top panel of Fig.~\ref{fig:linspec} shows the simulations and experimentally obtained linear absorption spectrum of Nile blue in ethanol\cite{Son2017UltrabroadbandDetection}. To allow easier comparison of the shape of the spectrum, the simulated spectrum has been shifted such that its absorption maximum is aligned with the experiment. The energy gap model was trained using TDDFT with the CAM-B3LYP exchange-correlation functional and 6-31G$^*$ basis set, which is known to systematically overestimate electronic excitation gaps of organic chromophores\cite{Laurent2013TD-DFTReview,Shao2019BenchmarkingBiochromophores} by 100s of meV, and in this case overestimates it by \numericalresult{0.32 eV} (the spectral maximum is 1.97 eV experimentally and 2.29 eV from simulation). One could improve the description of the position of the absorption maximum by using transfer learning to higher level wavefunction methods\cite{Chen2023ElucidatingStructure, Chen2023-MachineStates,Chen2023Data-EfficientAccuracy}, but since our primary focus here is the mechanisms by which solvent interactions give rise to the spectral shape, we apply a uniform \numericalresult{0.32 eV} shift of the linear and 2DES spectra to allow for easier comparison with experiment. With the shift applied, our simulated linear absorption spectrum yields good agreement with the experiment (Fig.~\ref{fig:linspec}) with a small underestimate of the width and high-frequency tail of the spectrum. The slightly narrower spectrum may be due to TDDFT with CAM-B3LYP underestimating the effect on the excitation energies upon H-bonding with the solvent, which has recently been demonstrated for the green fluorescent protein chromophore in water where EOM-CCSD was shown to correctly capture this effect\cite{Chen2023ElucidatingStructure}, or it may be due to non-Condon fluctuations, which have also recently been shown to produce additional broadening\cite{Wiethorn2023BeyondSimulations}.

To better understand the origins of the shape of the linear spectrum, we can analyze the spectral density, $J(\omega)$, which can be obtained from\cite{Mukamel1985FluorescenceEigenstates,Mukamel1995PrinciplesSpectroscopy,Zuehlsdorff2019OpticalApproaches} the time correlation of the energy gap fluctuations $\delta U$ relative to the mean electronic energy gap,   
\begin{equation}
    J(\omega) = \frac{\beta \omega}{2} \int_{-\infty}^\infty dt ~ e^{i\omega t} \langle \delta U(t) \delta U(0) \rangle.
\end{equation}
The spectral density encodes how strongly nuclear motions, including both those of the chromophore and its solvation environment, couple to the electronic excitation between the ground and first excited state, and thus provides information about how these motions modulate electronic energy transfer. The simulated spectral density is shown in the bottom panel of Fig.~\ref{fig:linspec} accompanied by the experimentally obtained surface-enhanced Raman scattering (SERS) spectrum of Nile blue in ethanol \cite{Auguie2012TinyAbsorptions}. The SERS spectrum and the simulated spectral density correspond to different observables and so the relative intensities of the peaks should not be directly compared but one generally expects to see similar features since they both are sensitive to motions that couple strongly to the electronic excitation. In particular, both have a sharp prominent peak at \numericalresult{$\sim$590 $\mathrm{cm}^{-1}$} that corresponds to a ring breathing mode of the central heteroatom-containing ring of Nile blue\cite{Lawless1992Excited-stateIntensities}. The other prominent experimentally observed feature, a 1640 $\mathrm{cm}^{-1}$ vibration primarily involving motion of the charged amine, is red-shifted and lower intensity than the corresponding highest frequency SERS peak. A normal mode analysis (SI Sec.~III) of the Nile blue molecule in vacuum using the same electronic structure method as that for training the ground state dynamics shows a set of four modes ranging from \numericalresult{1578 to 1636 $\cm$} that all contain contributions from the scissor motion of the -NH$_2^+$ (Fig.~\ref{fig:linspec} top, inset). These modes overlap closely with the peak in the simulated spectral density, indicating that the deviation from experiment is primarily a result of the level of electronic structure theory and not errors in the machine learning or effects from solvation. The difference in intensity reflects the different weighting SERS and spectral densities give to vibrational features with the low intensity in the simulated spectrum indicating that this vibrational mode is weakly coupled to the electronic transition.

\begin{figure}[h]
    \begin{center}
        \includegraphics[width=0.45\textwidth]{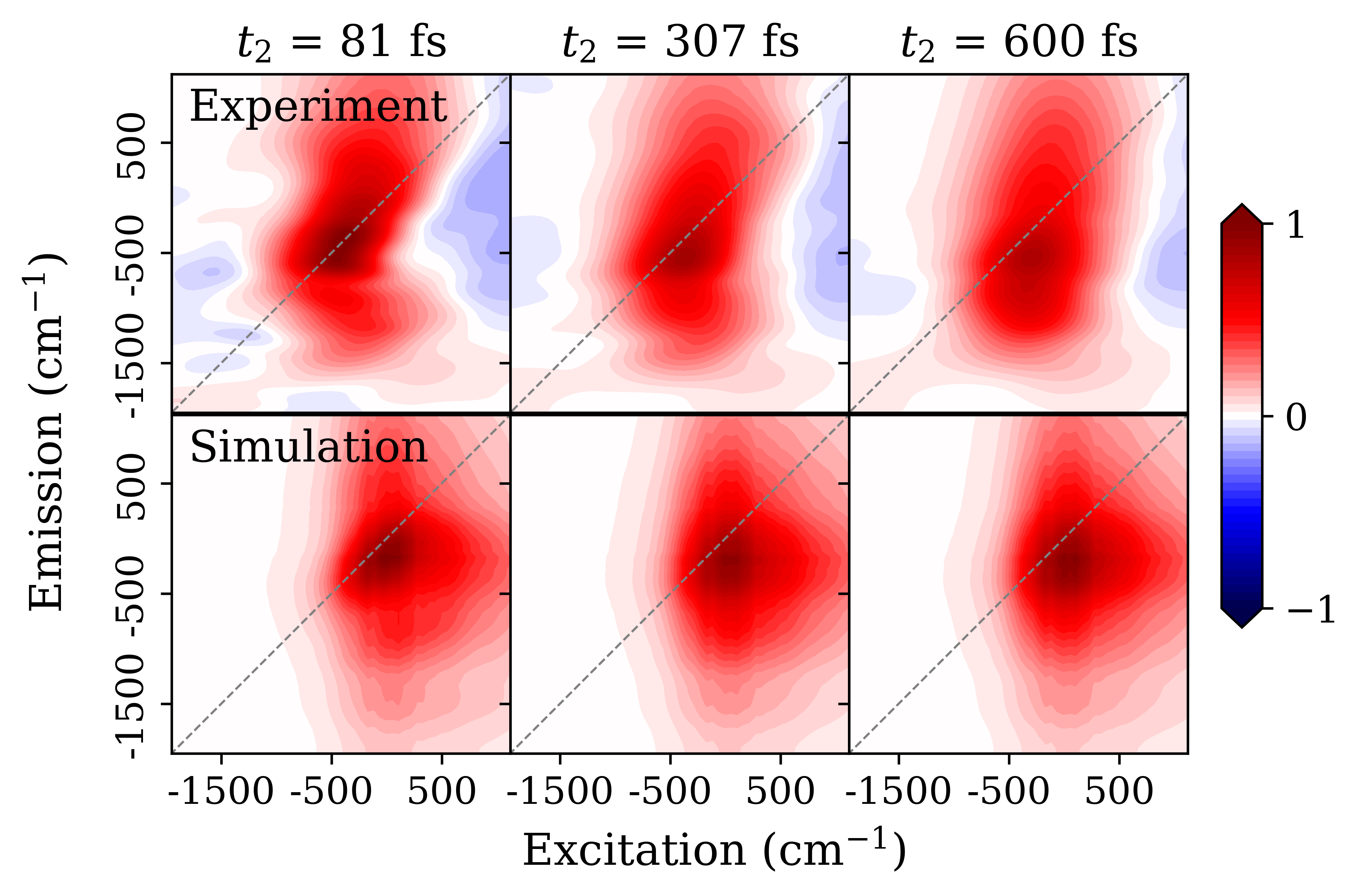}
    \end{center}
    \vspace{-5mm}
    \caption{Comparison between experimental and simulated 2DES spectra. Top: 2DES obtained from experiment\cite{Son2017UltrabroadbandDetection} at delay times of 81, 307, and 600~fs. Both excitation and emission frequencies are shifted by 1.97~eV as in Fig.~\ref{fig:linspec}. Bottom: The simulated spectra at the same delays. Both axes are shifted by 2.29~eV and have the experimental pulse spectral profile applied (SI Sec.~II).} 
    \label{fig:2des_direct}
\end{figure}

2DES spectra present a more strenuous test of simulation than linear absorption spectra owing to the richer details they encode. Fig.~\ref{fig:2des_direct} compares the 2DES spectra obtained from simulation to that from experiment. The simulated spectra are shifted by the same amount used for the linear spectra and we have multiplied the signal by the experimental pulse spectral profile to account for the window of frequencies observable in the experiment (SI Sec.~II). The features and general shape of the 2DES obtained from the experiment and simulation are in good agreement, except for the negative feature observed in the experimental signal to the right of the main peak. Such negative features can arise from excited state absorption (ESA) processes but transient absorption experiments suggest that the negative feature observed here is likely an artifact from the phasing process. \cite{Lewis2010Two-dimensionalTransition, Son2017UltrabroadbandDetection, Dostal2018DirectInteractions,Mewes2021BroadbandDyes}. The simulated signal does not have any negative intensity because our simulation only includes the ground and first excited states, so there are no higher electronic states to absorb into. While the negative features observed at higher excitation frequencies in the experimental 2DES likely arise from phasing artifacts, the experimental pump-probe spectra (SI Sec. IV) show negative intensity that arises from ESA above $\sim2000~\cm$. These pronounced negative intensity contributions from ESA should appear as negative intensity on the 2DES emission axis since the integral of the excitation axis onto it yields the time-dependent pump-probe spectra.\cite{Mukamel1995PrinciplesSpectroscopy,Jonas2003TWO-DIMENSIONALSPECTROSCOPY} However, these contributions lie above 2000 $\cm$ and hence they are not observed as they lie outside the range of observable frequencies due to the experiment's pulse spectral profile. Despite this, ESA still affects the experimental 2DES by shifting the position of the peak maximum to a lower frequency than that of the linear absorption spectrum. This shift in the 2DES maximum is not captured in the simulations that do not include ESA.

\begin{figure}[h]
    \begin{center}
        \includegraphics[width=0.45\textwidth]{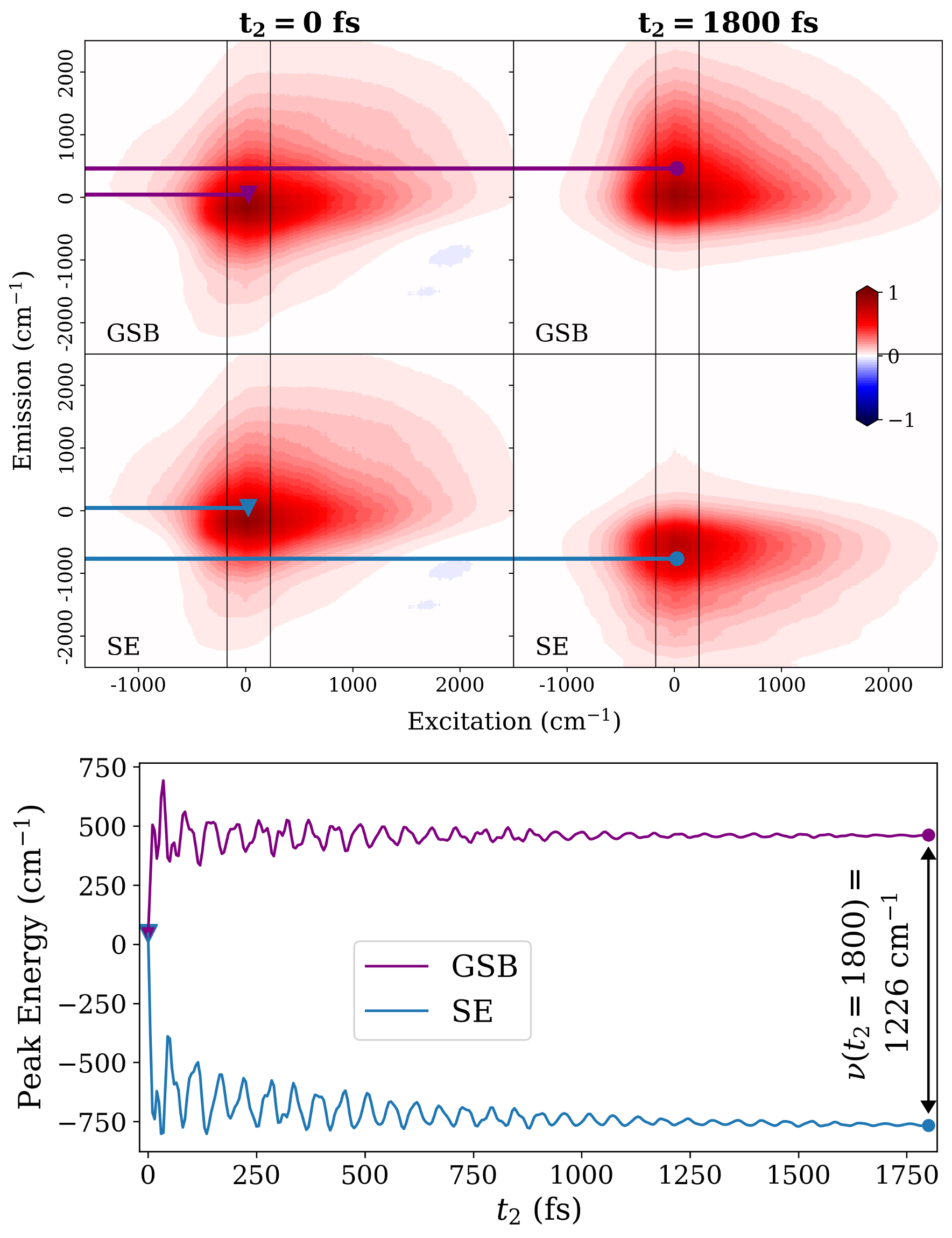}
    \end{center}
    \vspace{-5mm}
    \caption{Obtaining the Stokes shift and dynamic Stokes shift from our 2DES simulations. Top panels: The simulated 2DES spectra with the experimental pulse spectral profile applied is decomposed into its SE and GSB components at two time delays (left column: $t_2=0$~fs and right column: $t_2=1800$~fs). These components are identical at $t_2=0$~fs and then separate as the delay time is increased due to the stabilization of the excited state. The weighted means (blue and purple markers) are obtained by averaging the data between the black vertical lines and calculating the weighted mean emission frequency of the resulting distribution. Triangles and circles are used to show the position at $t_2=0~\mathrm{fs}$ and $t_2=1800~\mathrm{fs}$ respectively. Bottom panel: The position of the weighted means of the 2DES spectra for the SE and GSB  components as a function of the delay time. The separation of the two lines shown at each time is used to calculate the dynamic Stokes shift and the long time value (shown with double-headed arrow), \numericalresult{1226}~$\cm$, is the Stokes shift.} 
    \label{fig:se_gsb}
\end{figure}

Having established the accuracy of the simulated spectra, we can now use them to decompose the 2DES signal into its different contributions to uncover the role of solvent coupling to the excited state. The top panels of Fig.~\ref{fig:se_gsb} show the decomposition of the simulated signal into the contributions arising from the ground state bleach (GSB) and stimulated emission (SE) processes in which the chromophore population evolves on its ground or excited electronic state, respectively, during the $t_2$ delay. These processes are challenging to distinguish experimentally because both have the same wave vectors but are trivial to separate in the simulations since they arise from different terms in the response function\cite{Biswas2022CoherentSpectroscopies}. 

As the system undergoes longer $t_2$ delays, the SE peak shifts to lower energies since the chromophore and solvent rearrange to stabilize the excited electronic state. In contrast, the GSB peak arises from evolution on the ground state and shifts by $\sim$450 $\cm$ to higher energies in the first 50 fs (SI Sec.~V). By tracking the weighted mean position of the emission frequency of the spectra corresponding to the GSB and SE processes (purple and blue lines respectively in Fig.~\ref{fig:se_gsb}), the dynamics of the excited state stabilization by the motions of the chromophore and solvent can be examined. The bottom panel of Fig.~\ref{fig:se_gsb} shows the time evolution of the positions of the GSB and SE. As the $t_2$ time delay increases, the SE signal decreases in energy, leading to a larger separation between it and the GSB signal. The relaxation of the SE position can be decomposed into fast ($\sim$50 fs) and slow ($\sim$1 ps) components with the former corresponding to intramolecular relaxation within the existing solvation environment while the latter component corresponds to broader conformational rearrangements and relaxation of the solvation environment around Nile blue. The GSB and SE position of the weighted means oscillate with a $\sim$55 fs period corresponding to the prominent 590~$\cm$ ring breathing mode in the spectral density. The gap shown in Fig.~\ref{fig:se_gsb}, $\nu(t_2)$, can be used to calculate the dynamic Stokes shift,\cite{Yan1990FemtosecondPhases,Mukamel1995PrinciplesSpectroscopy} $S(t_2)$, according to
\begin{align}
    S(t_2) = \frac{\nu(t_2)-\nu(\infty)}{\nu(0)-\nu(\infty)}.
\end{align}
Within linear response, the dynamic Stokes shift is proportional to the energy gap time correlation function, with the corresponding Fourier transform used to construct the spectral density\cite{Mukamel1995PrinciplesSpectroscopy}.  

The difference between the GSB and SE in the long-time limit $\nu(t_2=1800~{\rm fs})$ in Fig.~\ref{fig:se_gsb} (marked with an arrow between the circles) can be measured as a shift of 1226 $\cm$. This shift is an underestimate since we applied the experimental pulse spectral profile to our simulated result, which narrows the range of emission frequencies that contribute significantly to the weighted mean. In particular, the SE peak extends to lower frequencies than the laser used in the experiment can fully capture. When simulated spectra are used without the experimental pulse spectral profile applied (SI Sec.~V), the predicted shift increases to \numericalresult{1524 $\cm$}. We note that peak maxima are often used to measure the splitting between the GSB and SE\cite{Lee2017UltrafastSpectroscopy}. However, due to the large asymmetry in the peaks, this approach leads to a $\sim$2-fold reduction of the shift to \numericalresult{529 $\cm$} with the pulse spectral profile applied and \numericalresult{623} $\cm$ without. 

The Stokes shift, the difference between the energy at which a molecule absorbs and fluoresces light, is defined using the maxima of the peaks. Experimentally, we measured this to be 1010 $\cm$ (SI Sec.~VI), consistent with previous values in the literature\cite{Baumann2001SolvationGlasses, Jose2006Benzophenoxazine-basedBiomolecules, GhanadzadehGilani2012ExcitedStudy} ranging from 891 to 956~$\cm$. However, due to the asymmetries in the absorption and florescence spectra, which arise from phenomena such as vibronic effects, the weighted mean frequency provides a better way to characterize the position of the distribution. Using the weighted mean of the experimental spectra gives a value of 1870 $\cm$  (SI Sec.~VI), which is significantly larger than the Stokes shift obtained from the spectral maxima. The difference in the positions between the weighted mean absorption and fluorescence spectra can be obtained from the simulations by reweighting the thermally sampled ground state configurations to account for their probabilities on the excited state (SI Sec.~VII).\cite{Zwanzig1954HighTemperatureGases, Tuckerman2010StatisticalSimulation, Ceriotti2012TheIntegration} Evaluating this from our simulations gives a shift of 1776 $\cm$, in good agreement with the experimental value obtained using the same approach (1870 $\cm$). 

The reorganization energy, $\lambda$, which characterizes the strength of the coupling between the electronic states and nuclear motions, is given by,
\begin{equation}
    \label{eq:reorganization_energy}
    \lambda = \frac{1}{\pi \hbar} \int_0^\infty \mathrm{d}\omega ~ \frac{J(\omega)}{\omega}
\end{equation}
where $J(\omega)$ is the spectral density. Based on our simulations, the reorganization energy is \numericalresult{924 $\cm$}. Under the second-order cumulant approximation, the reorganization energy is half the Stokes shift in the limit of Gaussian absorption and emission line shapes and equals half the difference between the average of the excitation energy in thermal equilibrium as measured on the ground- and excited-state potential energy surface, respectively.\cite{Mukamel1995PrinciplesSpectroscopy} From our simulations, using Eq.~\ref{eq:reorganization_energy} the reorganization energy is 924~$\cm$ and thus the Stokes shift is 1824 $\cm$, which is consistent with the value of 1776 $\cm$ obtained from the simulations by reweighting and the experimentally obtained difference between the weighted means of the absorption and fluorescence spectra of 1870 $\cm$. 

\begin{figure}[h]
    \begin{center}
        \includegraphics[width=0.4\textwidth]{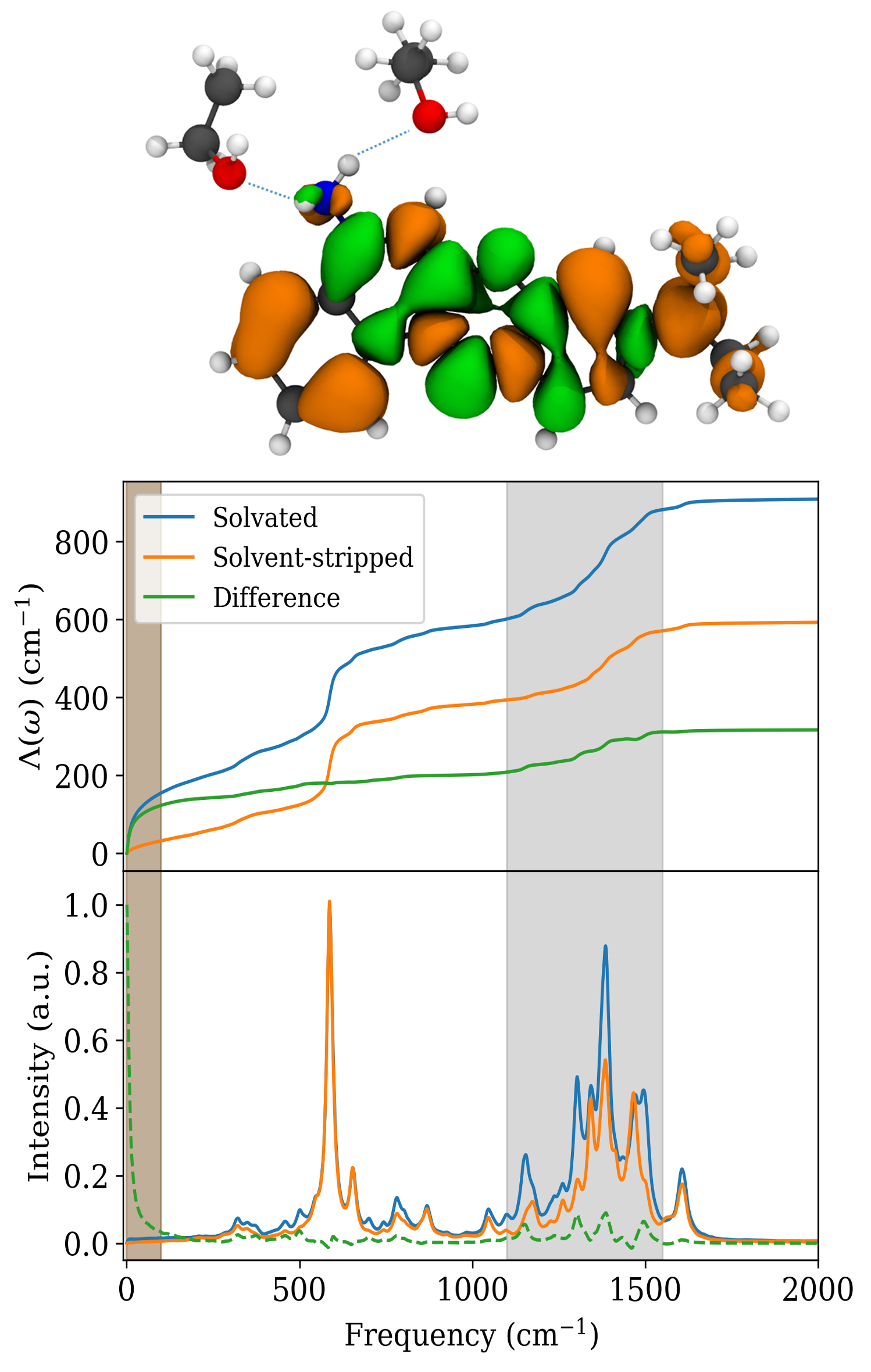}
    \end{center}
    \vspace{-5mm}
    \caption{Analysis of the changes to the reorganization energy and spectral density upon stripping solvent. Top: Difference between the ground and S$_1$ excited state electron density of Nile blue shown with two ethanol molecules hydrogen bonded to the primary amine (-NH$_2$) group. The difference density reveals significant changes to the electron density across the entire Nile blue molecule. Middle: $\Lambda(\omega)$ for the solvated (blue) and solvent-stripped (orange) simulations along with the difference between them (green). The values at $\omega = 2000~\cm$ for the solvated and solvent-stripped simulations (\numericalresult{890} $\cm$ and \numericalresult{592 $\cm$} respectively) are their reorganization energies. These reorganization energies have significant contributions from low frequency motion below 100 $\cm$ (brown shaded), a sharp increase just below 600 $\cm$, and in the 1100-1550 $\cm$ region (shaded gray). However, the difference arises primarily from the two shaded regions only. Bottom: The spectral densities of the solvated (blue) and solvent-stripped (orange) simulations. The solvated spectral density is normalized such that the maximum value is 1 and the solvent-stripped spectral density is normalized with the same factor. The derivative of the difference in $\Lambda(\omega)$, $\frac{d}{d\omega}\Delta\Lambda(\omega)$ is also plotted (dashed green) to highlight the largest contributions to the deviation between solvated and solvent-stripped simulations.} 
    \label{fig:strippedsolvent}
\end{figure}

To isolate the microscopic origin of the stabilization of the excited state, we now consider the effect of removing the electronic contribution of the solvent from the simulated predictions and its effect on the reorganization energy\cite{Lu2021TheSpectroscopy,Myers2024AxialMethanol}. To do this, we took the training configurations used previously, which were generated from the ground state dynamics of Nile blue in ethanol, and removed (stripped) all the solvent from those configurations. We then ran excited state electronic structure calculations on these solvent-stripped configurations, which were then used to train an EqT electronic energy gap model. This model was then used to compute the electronic energy gaps for the whole Nile blue in ethanol trajectory, considering only the chromophore configurations. The solvent-stripped spectral density (Fig.~\ref{fig:strippedsolvent}) was then computed from these energy gaps. The resulting spectral density thus retains the same nuclear motions of the chromophore and, therefore, dynamic and structural changes that arise from it being in solution, but does not include the electronic effects of the solvent on modulating the electronic energy gap between the ground and excited state.

The changes upon stripping the solvent are shown in Fig.~\ref{fig:strippedsolvent} where we computed the following quantity,
\begin{equation}
    \Lambda(\omega) = \frac{1}{\pi \hbar} \int_0^\omega \mathrm{d}\omega' ~ \frac{J(\omega') }{\omega'}.
\end{equation}
By comparing the above expression to Eq.~\ref{eq:reorganization_energy} one can see that this quantity is simply the contribution to the reorganization energy arising from the spectral density up to frequency $\omega$ and hence $\Lambda(\omega\to \infty)=\lambda$. The top panel of Fig~\ref{fig:strippedsolvent} shows $\Lambda(\omega)$ for the solvated and solvent-stripped simulations as well as their difference. From this, one can see that the lower overall reorganization energy in the solvent-stripped simulation \numericalresult{(592 $\cm$)}, in contrast to the fully solvated value of \numericalresult{890~$\cm$}, predominantly arises from two frequency regions, \numericalresult{0-100 $\cm$} and \numericalresult{1100-1550~$\cm$} (shaded in brown and gray respectively in Fig.~\ref{fig:strippedsolvent}) with the former accounting for \numericalresult{39$\%$} and the latter \numericalresult{33$\%$} of the change in the reorganization energy, with the much wider frequency region between them accumulating the remaining \numericalresult{28$\%$} change. The lower frequency region (shaded brown) can be physically interpreted as solvent motions along with the collective flexing of the Nile blue aromatic structure, while the higher frequency region (shaded gray) arises from large internal motions of the molecule involving C-C stretches within the aromatic rings and C-H bends in the diethyl-amine functional group. The bottom panel of Fig.~\ref{fig:strippedsolvent} shows the spectral densities from the solvated and solvent-stripped simulations along with the derivative of the difference, 
\begin{equation}
    \frac{d\Delta\Lambda(\omega)}{d\omega} = \frac{J_{solvated}(\omega) - J_{stripped}(\omega)}{\omega}
\end{equation}
where $\Delta\Lambda(\omega)=\Lambda_{solvated}(\omega)-\Lambda_{stripped}(\omega)$, $\Lambda_{solvated}(\omega)$ and $\Lambda_{stripped}(\omega)$ denote the $\Lambda(\omega)$ from the solvated and solvent-stripped simulations respectively. $J_{solvated}(\omega)$ and $J_{stripped}(\omega)$ denote the corresponding spectral densities. From this one can see that the intense ring breathing mode of the central heteroatom-containing ring (\numericalresult{587 $\cm$}) makes little contribution to the change in the reorganization energy upon stripping the solvent, indicating that the solvent-induced electronic structure changes do not mediate the coupling of this vibration to the electronic transition.

In contrast, there is an increase in the reorganization energy due to the presence of solvent for some of the higher-frequency peaks in the spectral density (gray shaded region). To better understand why the solvent more strongly couples to certain nuclear motions, we analyzed the electron density difference upon excitation and the extent of H-bonding between the chromophore and ethanol solvent molecules. 

The difference between the ground and excited-state electron density (Fig \ref{fig:strippedsolvent}, top) shows that the most notable changes occur in the ring systems. However, the H-bond donating primary amine, although not exhibiting a large difference in electron density itself, provides a pathway to couple solvation effects to electron density changes in the ring system. This amine shows significant H-bonding in the molecular dynamics simulations, acting as an H-bond donor to the solvent ethanol molecules 98.5\% of the time (with two H-bonds to ethanol shown in Fig \ref{fig:strippedsolvent}, top), whereas the chromophore ring oxygen or nitrogen atoms act as H-bond acceptors at most 0.8\% of the time and the nitrogen in the sterically hindered tertiary amine acts as an H-bond acceptor just 0.1\% of the time (SI Sec.~VIII). Thus, the high frequency modes that contribute to the change in the reorganization energy due to the solvent are those that involve ring modes that can receive or donate electron density via the primary amine. We note, however, that the mode at 1640 $\cm$ itself does not change much in intensity upon solvation since this mode predominantly involves the amine with less contribution from the ring systems (SI Sec.~III). In addition, the central ring breathing mode at \numericalresult{587 $\cm$} (SI Sec.~III), although possessing a large contribution to the change in the electron density upon excitation and hence appearing intensely in the spectral density, does not exhibit sensitivity to solvent and hence does not contribute to the change in the reorganization energy between the solvated and solvent-stripped simulations. These observations highlight the importance of considering both the locations of the changes to the electron density upon moving from the ground to the excited state and also how H-bonding sites interact with that electron density.  Both of these factors affect the reorganization energy and the overall ability of solvent to control relaxation on the excited state surface.  

In summary, we have shown that one can construct ML potential energy surfaces using the EqT architecture capable of generating linear and 2DES spectra for a chromophore in its condensed phase chemical environment with orders of magnitude fewer ground and excited state electronic structure calculations than would be needed to perform these calculations directly using {\it ab initio} molecular dynamics. By applying this approach to Nile blue in ethanol and comparing it to recent experiments, we show that we can obtain good agreement with both the shape of the linear, pump-probe, and multidimensional spectra by combining TDDFT electronic structure with a second-order cumulant expansion in the electronic energy gap fluctuations treatment of the spectra. By rigorously breaking down the simulated spectra into signals arising from stimulated emission and the ground-state bleach, we assessed the dynamic Stokes shift and the emergence of the shifts arising from the long-time stabilization of the electronic excited state due to nuclear motion of the chromophore and solvent. Finally, by creating an electronic energy gap surface trained on solvent-stripped electronic structure calculations, we have shown how one can break down the reorganization energy into frequency-dependent components that indicate the key regions of nuclear motions of the chromophore and solvent that couple to the electronic transition. We believe that the ability to efficiently and accurately simulate the complex interactions that contribute to multidimensional spectra enabled by these developments will lead to deeper synergies between experiment and theory going forward. 2DES has been widely applied to molecular, materials, and biological systems that rely on coupled electronic and nuclear dynamics, but specific molecular assignments to features and phenomena within these spectra has often been intractable. Through this advance, we can now simulate these measurements, enabling interpretation through the lens of ground-up theories.

\section*{Supporting Information}
Additional details of the experimental methods, electronic structure calculations, training of machine learning models, molecular dynamics simulations, assessment of ML-based computational cost acceleration, normal modes that contribute to the spectral density of Nile blue in ethanol, comparison between experimental and simulated pump-probe spectra, assessment of the GSB and SE components of the simulated 2DES spectra without the laser profile, reweighting procedure to obtain the Stoke shift, and hydrogen bonding sites of Nile blue (PDF)

Data sets, EqT models, and input files to train the ground state and electronic excitation energy gap EqT models (10.5281/zenodo.15376273)

\section*{Acknowledgments}
This work was funded by the U.S. Department of Energy, Office of Science, Office of Basic Energy Sciences (DE-SC0020203 to C.M.I. and T.E.M.) and the National Science Foundation Grant No. CHE-2154291 (to T.E.M.). M.S., A.L., and G.S.S.-C. were funded by the U.S. Department of Energy, Office of Science, Office of Basic Energy Sciences, Division of Chemical Sciences, Geosciences, and Biosciences under Award \#DE-SC0018097 to G.S.S.-C. M.S.C. acknowledges support from the Simons Center for Computational Physical Chemistry at New York University. This research used resources of the National Energy Research Scientific Computing Center (NERSC), a U.S. Department of Energy Office of Science User Facility located at Lawrence Berkeley National Laboratory, operated under Contract No. DE-AC02-05CH11231 using NERSC Awards BES-ERCAP0019987 and BES-ERCAP0023755. A.M.C.~was funded by an Early Career Award in the CPIMS program in the Chemical Sciences, Geosciences, and Biosciences Division of the Office of Basic Energy Sciences of the U.S. Department of Energy under Award No.~DE-SC0024154. Some of the computing for this project was performed on the Sherlock cluster. We would like to thank Stanford University and the Stanford Research Computing Center for providing computational resources and support that contributed to these research results.

\bibliography{references}
\end{document}


\title{Supporting Information: Two-dimensional electronic spectroscopy in the condensed phase using equivariant transformer accelerated molecular dynamics simulations}

\author{Joseph Kelly}
\affiliation{Department of Chemistry, Stanford University, Stanford, California, 94305, USA}

\author{Frank Hu}
\affiliation{Department of Chemistry, Stanford University, Stanford, California, 94305, USA}

\author{Arianna Damiani}
\affiliation{Department of Chemistry, Stanford University, Stanford, California, 94305, USA}

\author{Michael S. Chen}
\affiliation{Simons Center for Computational Physical Chemistry, Department of Chemistry, New York University, New York, New York 10003, United States}

\author{Andrew Snider}
\affiliation{Chemistry and Chemical Biology, University of California Merced, Merced, California 95343, USA}

\author{Minjung Son}
\affiliation{Department of Chemistry, Boston University, Boston, Massachusetts 02215, USA}

\author{Angela Lee}
\affiliation{Department of Chemistry, Massachusetts Institute of Technology, Cambridge, Massachusetts 02139, USA}

\author{Prachi Gupta}
\affiliation{Chemistry and Chemical Biology, University of California Merced, Merced, California 95343, USA}

\author{Andr\'es Montoya-Castillo}
\affiliation{Department of Chemistry, University of Colorado, Boulder, Boulder, Colorado, 80309, USA}

\author{Tim J. Zuehlsdorff}
 \affiliation{Department of Chemistry, Oregon State University, Corvallis, Oregon 97331, USA}

\author{Gabriela S. Schlau-Cohen}
\email{gssc@mit.edu}
\affiliation{Department of Chemistry, Massachusetts Institute of Technology, Cambridge, Massachusetts 02139, USA}

\author{Christine M. Isborn}
\email{cisborn@ucmerced.edu}
\affiliation{Chemistry and Chemical Biology, University of California Merced, Merced, California 95343, USA}

\author{Thomas E. Markland}
\email{tmarkland@stanford.edu}
\affiliation{Department of Chemistry, Stanford University, Stanford, California, 94305, USA}

\date{\today}

\maketitle
\normalsize
\tableofcontents
\newpage

\section{Methods}
\subsection{Experimental}
Steady-state measurements were performed with Nile Blue A perchlorate (Millipore Sigma) dissolved in spectroscopic-grade ethanol (Millipore Sigma) with an OD of ~1.5 at 633 nm using a 0.5 mm cuvette. For fluorescence measurements, the sample was excited at 550 nm. TG-FROG was used to characterize the pulse duration (13 fs) and the spectral profile of the compressed pulse (SI Fig. \ref{fig:si_laserprofile}, right). 2DES experiments were performed on a fully noncollinear BOXCARs setup. Full details can be found in Son, et. al.\cite{Son2017UltrabroadbandDetection} Nile Blue A perchlorate measurements were performed with the same sample optical density and laser pulse energy as in Son, et. al., although with coherence times scanned over $t_1=-80~\mathrm{to}~80$ fs with 0.4 fs steps and waiting times sampled at $t_2$ = 0 fs and 68 – 600 fs in 13 fs steps. Spectra were phased using the projection slice theorem\cite{Jonas2003TWO-DIMENSIONALSPECTROSCOPY}. 

\subsection{Ground State Equivariant Transformer Training and Dynamics}
\label{sec:methods_grndEqT}
To obtain the ground state trajectory of Nile blue in ethanol, we trained a machine-learned potential based on an equivariant transformer (EqT) architecture\cite{Vaswani2017AttentionNeed,Tholke2021EquivariantPotentials} using TorchMD-NET\cite{Pelaez2024TorchMD-NetSimulations,GitHubPotentials} version 2.1.0. The EqT generates a learned feature vector which captures both the properties of an atom and its neighborhood and encodes permutational, rotational, and translational symmetries. It then iteratively updates this feature vector using an attention mechanism that weights interatomic interactions by their distances, and predicts properties via an output network operating on these feature vectors. 
To build a training set for the EqT model, we performed multiple iterations of training EqT models and running dynamics. Table~\ref{table:curation_iterations} summarizes each iteration of the data curation and the model architectures that were trained on those configurations to produce the trajectories used for the next iteration. The model architectures referenced in Table~\ref{table:curation_iterations} are summarized in Table~\ref{table:model_architectures}. To mitigate overfitting, we employed early-stopping by assessing the loss of a validation set at each epoch. The loss was an RMSE error for energy gaps of each configuration and force components in the x, y, and z dimension for each atom in the configuration with the energy and force components weighted with ratios described in Table~\ref{table:model_architectures}. Models were trained with the Adam optimizer, a batch size of 8, a learning rate of 0.001, and learning rate reduction on plateau scheduler using a reduction factor of 0.9, and patience of 5 epochs\cite{KellyDatasetsSimulations}. All results in the main text were obtained from a 1-ns trajectory calculated using the EqT Final model. 
All molecular dynamics trajectories were performed using a cubic box with side length 25.75 Å, 175 ethanol molecules, and the positively charged Nile blue chromophore. The initial molecular dynamics trajectory was performed using density functional tight binding (DFTB) theory \cite{Porezag1994ConstructionCarbon, Seifert1996CalculationsScheme} with the i-PI program\cite{Kapil2019I-PISimulations} using the CP2K program\cite{Hutter2014Cp2k:Systems} as the force engine. The 3ob parameter set\cite{Gaus2013ParametrizationMolecules} was used with dispersion forces included via a Lennard-Jones potential\cite{Zhechkov2005AnBinding} with parameters taken from the Universal Force Field\cite{Cundary1992UFFSimulations}. The trajectory was performed with a time step of 0.5 fs in the NVT ensemble using a stochastic velocity rescaling thermostat\cite{Bussi2007CanonicalRescaling} with a 1 ps time constant at 300 K. \revision{DFTB was used as a computationally efficient method to sample an initial ensemble of decorrelated configurations on which DFT calculations were performed to generate training data for the machine learned potentials.} All EqT-based molecular dynamics simulations were performed using OpenMM\cite{Eastman2024OpenMMPotentials} and the openmm-torch plugin with a time step of 0.5 fs in the NVT ensemble using a local Langevin thermostat\cite{Schneider1978Molecular-dynamicsTransitions} with a 1 ps time constant at 300 K. The DFT calculations were performed using CP2K\cite{Hutter2014Cp2k:Systems} with the revPBE exchange-correlation functional\cite{Perdew1996GeneralizedSimple,Zhang1998CommentSimple}, D3 dispersion \cite{Grimme2010AH-Pu} and the TZV2P-GTH basis set. The functional revPBE has been shown to be effective at capturing structural and spectroscopic properties of condensed phase systems\cite{Goerigk2011AInteractions,Remsing2014TheModels,Bankura2014StructureDynamics,Skinner2016TheSimulation,Marsalek2017QuantumEffects}. 

\begingroup
\setlength{\tabcolsep}{10pt}
\begin{table}[!ht]
    \centering
    \caption{Summary of the data curation iterations used for generating a training set for the EqT model. For each iteration, the breakdown of the configurations used for training is provided in parentheses if the set used for training was a combination of previous sets of configurations. \revision{Iteration 3 removed 38 converged configurations that were outliers from the original set: defined as having energies more than 3 standard deviations above the mean of the full set.}}
    \label{table:curation_iterations}
    \large
    \resizebox{\columnwidth}{!}{\begin{tabular}{lp{9cm}llll}
        \toprule
        \textbf{Iteration} & \textbf{Starting trajectory} & \textbf{configurations sampled} & \textbf{configurations converged} & \textbf{configurations for training} & \textbf{Model architecture(s) trained} \\ \midrule
        1 & 50 ps DFTB, 3ob-3-1 parameters & 250 (Sampled randomly) & 249 & 249 & EqT 1 \\
        2 & 32 ps trajectory from EqT 1 & 61 (Equidistant sampling from 10 - 32 ps) & 61 & 310 (249 + 61) & EqT 2 \\ 
        3 & 100 ps trajectory from EqT 2 & 2000 (Sampled starting at 20 ps with 40 fs spacing) & 1984 & 1946 & EqT 3, EqT 4, EqT 5, EqT Ensemble\\
        4 & 100~ps EqT 3 trajectory, 1~ns EqT 4 trajectory, 300~ps EqT 5 trajectory, 700~ps EqT Ensemble trajectory & 603 (Sampled near topology failures in each trajectory) & 600 & 2546 (1946 + 600) & EqT Final \\ \midrule
    \end{tabular}}
\end{table}
\endgroup

\begingroup
\setlength{\tabcolsep}{10pt}
\begin{table}[!ht]
    \centering
    \caption{Summary of EqT model architectures trained during the data curation process. Only the parameters listed in this table were varied, while all other parameters remained constant across different training runs.}
    \label{table:model_architectures}
    \large
    \resizebox{\columnwidth}{!}{\begin{tabular}{llllllll}
        \toprule
        \textbf{Model architecture} & \textbf{Energy weight} & \textbf{Force weight} & \textbf{Number of heads} & \textbf{Embedding dimension} & \textbf{Number of layers} & \textbf{Number of RBFs} & \textbf{Reduction operation} \\ \midrule
        EqT 1, EqT 2 & 0 & 1 & 2 & 64 & 1 & 32 & mean \\  
        EqT 3, EqT 4, EqT Ensemble, EqT Final & 1 & 10000 & 4 & 64 & 2 & 64 & mean \\ 
        EqT 5 & 1 & 10000 & 4 & 64 & 2 & 64 & add \\ \midrule
    \end{tabular}}
\end{table}
\endgroup

We note that a potentially even more compact training set could be achieved by using query by committee approaches \cite{Schran2020CommitteeLearning, Stolte2024RandomWater} but given the rapid convergence of the EqT architecture with just a few thousand training points we did not pursue those approaches here.

\subsection{Excitation Energy Calculations}
\label{sec:tddft}
The excitation energies, ground state gradients, and excited state gradients used to train the EqT energy gap model were evaluated with Time-Dependent Density Functional Theory (TDDFT) using the range-separated hybrid CAM-B3LYP \cite{Yanai2004ACAM-B3LYP} functional and 6-31G* \cite{Frisch1984SelfconsistentSets} basis set. This range-separate hybrid functional with long-range exact exchange was used for the excited state calculations to more accurately describe states with charge transfer character.\cite{Zuehlsdorff2020InfluenceDensity,Lu2021TheSpectroscopy} All calculations were run using TeraChem v1.9 \cite{Seritan2021TeraChem:Dynamics} on NVIDIA A100 GPUs. The training set of \numericalresult{220} configurations was obtained by randomly sampling along the ground state EqT trajectory. MD configurations were converted from periodic boundary conditions to cluster format for the TDDFT calculations. 
To ensure that the Nile blue molecule was well solvated and that conversion from a periodic system to a cluster did not leave molecular fragments, each configuration was redefined so the center of mass of the Nile blue molecule was at the center of the simulation box and all solvent molecules with at least 1 atom less than 10 Å from the center of mass of Nile blue were retained in whole. 
Benchmarking the effect of varying the QM solvent shell radius showed minimal change in the excitation energy values past 10 Å. Therefore, this distance was used to establish the final quantum mechanical region for the TDDFT calculations, which included the Nile blue chromophore and an average of 118 ethanol solvent molecules. All further solvent was replaced with an implicit Polarizable Continuum Model (PCM) within the COSMO \cite{Klamt1993COSMO:Gradient} framework using a dielectric constant $\epsilon$ = 24.5 for ethanol and the solute cavity generated by scaling the individual atomic radii by 1.5\cite{Provorse2016ConvergenceModels}.

To train the stripped-solvent electronic energy gap model, 598 configurations were randomly selected from the DFTB trajectory such that each sample was at least 50 fs from every other sample. All the solvent was removed from these configurations and TDDFT calculations were completed to obtain S$_0$ to S$_1$ excitation energies and gradients using the same method described above except no PCM was added around the Nile blue molecule.

\subsection{Excitation Energy Gap Equivariant Transformer}
\label{si_sec:excitation_ML}
The electronic excitation energy gaps from the S$_0$ to S$_1$ state and the gradients of this difference with respect to nuclear positions obtained from the above TDDFT calculations were used to train a set of EqT electronic energy gap models\cite{Vaswani2017AttentionNeed,Tholke2021EquivariantPotentials,Pelaez2024TorchMD-NetSimulations}. While the EqT models were trained using truncated clusters with, on average, 118 ethanol molecules (SI Sec. \ref{sec:tddft}), the final predictions were obtained using all the solvent from the dynamics. Each configuration was independently wrapped so that the Nile blue center of mass aligned with the center of the simulation box and all ethanol molecules that crossed the periodic boundary were randomly assigned to reside entirely on one side or the other. Each model had 4 heads, 2 layers, 64 radial basis functions (RBFs), an embedding dimension of 64, and an atom-wise sum reduction. To mitigate overfitting, we employed early-stopping by assessing the loss of a validation set at each epoch. Three models were trained using the energy gaps and gradients with 45, 90, and 180 configurations in the training set, 50, 10, and 20 configurations in the test set, and 5, 10, and 20 configurations in the validation set respectively. The loss was an RMSE error for energy gaps of each configuration and gradient components in the x, y, and z dimension for each atom in the configuration with the energy and gradient components weighted with a ratio of 1:1000. Models were trained with the Adam optimizer, a batch size of 1, a learning rate of 0.0004, and learning rate reduction on plateau scheduler using a reduction factor of 0.8, and patience of 15 epochs\cite{KellyDatasetsSimulations}. The same architecture and training procedure was used to construct an EqT model from the solvent-stripped TDDFT calculations with 477, 60, 60 configurations used in the train, validation, and test sets respectively. 
\newpage

\subsection{\revision{Computational costs and estimated acceleration from machine learning}}

\revision{To provide an assessment of the computational times (costs) and hence the acceleration over using electronic structure methods achieved by training machine-learned potentials for the ground state (Table~\ref{table:ground_state_speeds}) and electronic excitation energy gap (Table~\ref{table:energy_gap_speeds}) of solvated Nile blue in ethanol, here we provide the details of the computational time required for the electronic structure calculations and the model training and execution of the machine-learned potentials.}

\revision{In summary, the EqT models achieve a 
41,000 fold and 650,000 fold speedup relative to DFT and TDDFT computational cost per configuration respectively. Each EqT model requires an upfront cost to generate training data and train the model. The training of each EqT model itself requires $\sim$24~hours using 1 NVIDIA A100 GPU and required $\sim$2500 ground state DFT calculations and $\sim$100 TDDFT calculations which take $\sim$1250 CPU node-hours and $\sim$75 GPU node-hours respectively.}

\subsubsection{\revision{Ground state computational costs}}

\revision{Table \ref{table:ground_state_speeds} summarizes the cost per configuration for the DFT, DFTB, and ML model calculations for the ground state of Nile blue in 175 ethanol molecules (1619 atoms). From this one can see that the ML (EqT) model on a NVIDIA A100 GPU is 41,000 fold faster to evaluate than DFT and 130 fold faster than DFTB performed on 128 CPU cores.} 

\revision{The DFT and DFTB timings were performed using CP2K on 1 CPU node of DOE NERSC's Perlmutter supercomputer, which contains 2 AMD EPYC 7763's (128 cores). While the DFTB dynamics we used to generate the configurations for the machine learning training used i-PI for the evolution, for consistency of platforms, the DFTB timings were performed directly in CP2K. For the EqT models of the ground state, the timings reported were peformed using 1 NVIDIA A100 (40 GB) on DOE NERSC's Perlmutter supercomputer using TorchMD-NET.}

\begingroup
\setlength{\tabcolsep}{10pt}
\begin{table}[!ht]
    \centering
    \caption{\revision{Summary of the computational costs of the methods used in the ground state dynamics on 128 CPU cores for the DFT and DFTB calculations and a single NVIDIA A100 GPU for the ML (EqT) model. For the DFT calculations the cost per configuration is an average of each single-point calculation. The cost per configuration for the DFTB and the ground state EqT model were calculated by dividing the total time to simulate a trajectory by the number of time steps completed.}}
    \label{table:ground_state_speeds}
    \large
    \resizebox{.4\textwidth}{!}{\begin{tabular}{cc}
        \toprule
        \revision{\textbf{Method}} & \textbf{\revision{Cost per Configuration (sec)}} \\ 
        \midrule
        \revision{DFT} & \revision{1800}  \\
        \revision{DFTB} & \revision{5.6}  \\   
        \revision{EqT} & \revision{0.043} \\
        \midrule
    \end{tabular}}
\end{table}
\endgroup

\subsubsection{\revision{Excited state computational costs}}

\revision{Table \ref{table:energy_gap_speeds} summarizes the cost per configuration for the TDDFT with and without force evaluation and ML model calculations for the electronic excitation energy gap of Nile blue in ethanol. From this one can see that the ML model is 650,000 fold faster to evaluate than TDDFT with forces and 400,000 fold faster than energy only TDDFT calculations with all calculations performed on four NVIDIA A100 GPUs.}

\revision{The TDDFT calculations were performed in Terachem on 1 GPU node of DOE NERSC's Perlmutter supercomputer containing 4 NVIDIA A100's (40 GB) for Nile blue in ethanol molecules per configuration. For the EqT models of the electronic excitation energy gap, the training timings are reported on one NVIDIA A100 (40 GB) on DOE NERSC's Perlmutter supercomputer. The inference timings to predict excitation energy gaps for each configuration are reported using four GPUs to match the resources used on the TDDFT calculations.} 

\begingroup
\setlength{\tabcolsep}{10pt}
\begin{table}[!ht]
    \centering
    \caption{\revision{Summary of the computational costs of the methods used for the excitation energy gap using four NVIDIA A100 GPUs of the solvated Nile blue chromophore. Note that the excited state EqT model timings are an order of magnitude faster than for the ground state due to the use of 4 GPUs and that for the energy gap predictions, they can be performed in batches of 2.}}
    \label{table:energy_gap_speeds}
    \large
    \resizebox{.45\textwidth}{!}
    {\begin{tabular}{cc}
        \toprule
        \revision{\textbf{Method}} & \textbf{\revision{Cost per Configuration (sec)}} \\ 
        \midrule
        \revision{TDDFT with forces} & \revision{2700} \\  
        \revision{TDDFT energy only} & \revision{1680} \\ 
        \revision{EqT} & \revision{0.0041} \\
        \midrule
    \end{tabular}}
\end{table}
\endgroup

\newpage
\section{Experimental pulse spectral profile for 2DES}
\begin{figure}
    \centering\includegraphics[width=0.8\linewidth]{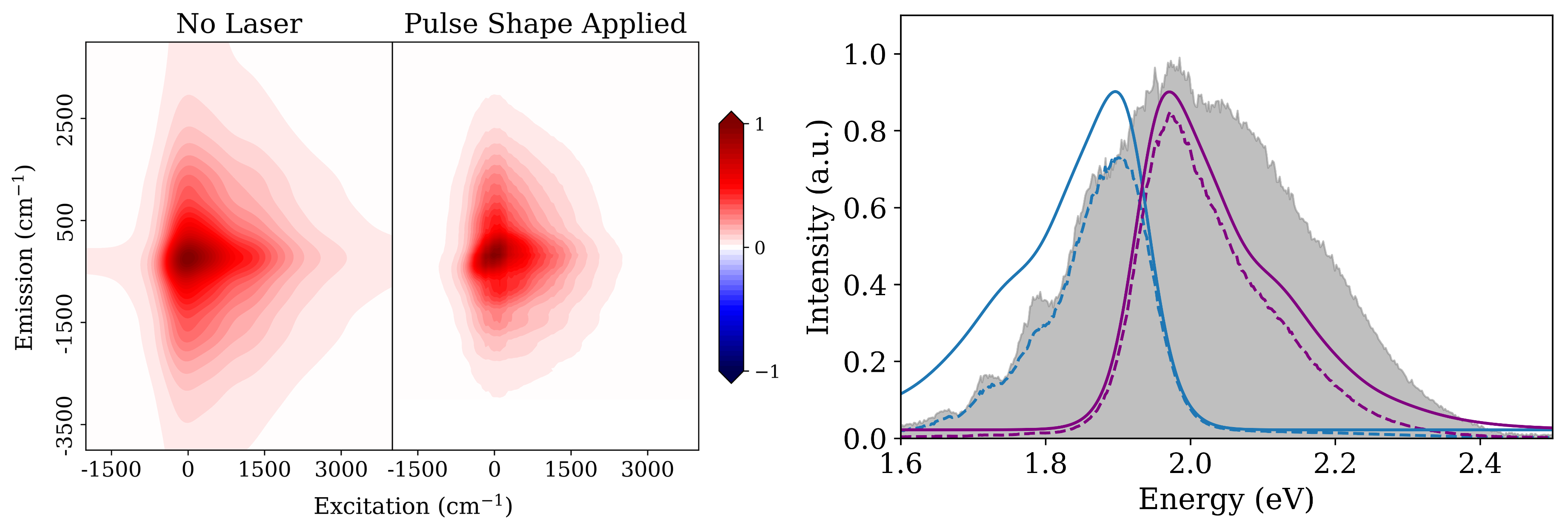}
    \caption{The experimental pulse spectral profile applied to the 2DES simulated data. Left panel: the simulated 2DES spectra at $t_2=600~\mathrm{fs}$ before and after applying the experimental pulse spectral profile. Right: The GSB (purple) and SE (blue) contributions to the simulated 2DES for $t_2=600~\mathrm{fs}$ and $\omega_3 = 0~\cm$ with the experimental pulse spectral profile applied (dashed) or not applied (solid). The spectral profile of the compressed laser pulse as determined through TG-FROG (gray shaded) is normalized to a maximum value of 1.}
    \label{fig:si_laserprofile}
\end{figure}

Experimental 2DES measurements probe the frequency range within the pulse spectral profile that was used. To allow for a direct comparison between the experiment and simulation, we applied the experimental pulse spectral profile of the compressed laser pulse as determined through TG-FROG to the simulated 2DES spectra. The right panel of SI Fig. \ref{fig:si_laserprofile} shows the intensity of the pulse spectral profile in gray. This was applied to the simulated 2DES spectra by multiplying the intensity along the $\omega_1$ axis and square root of the intensity along the $\omega_3$ to account for the use of two laser pump pulses to generate the signal along $\omega_1$ and one probe pulse to extract the information along $\omega_3$. 
The left panel of SI Fig. \ref{fig:si_laserprofile} shows the 2DES spectra at $t_2=600~\mathrm{fs}$ with and without the pulse spectral profile applied. From this one can see the effect of applying the pulse spectral profile is to narrow the spectrum without changing the overall shape or features in the spectra. The right panel of SI Fig. \ref{fig:si_laserprofile} shows the ground state bleach (GSB) component (purple) and stimulated emission (SE) component (blue) of the 2DES spectrum at $t_2=600~\mathrm{fs}$ and $\omega_3 = 0~\cm$. This shows that the GSB component is located within the pulse spectral profile and is therefore minimally affected by applying the pulse spectral profile. However, the SE component is lower in energy resulting in the pulse spectral profile truncating it at lower frequencies.
\newpage 

\section{Analysis of Normal Modes that Contribute to the Spectral Density}
\begin{figure}
    \centering
    \begin{subfigure}{0.7\textwidth}
        \centering
        \includegraphics[width=0.6\textwidth]{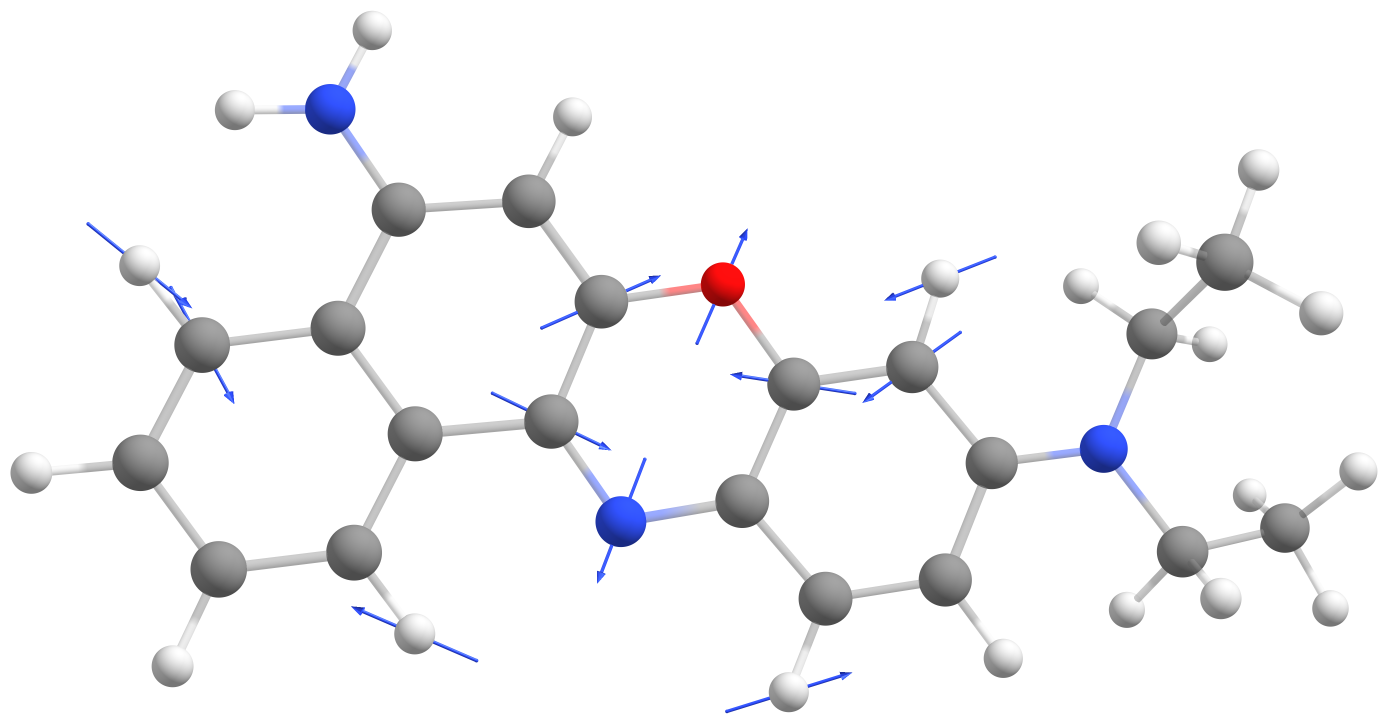}
        \caption{583 cm$^{-1}$}
        \label{fig:nb_nm_583}
    \end{subfigure}
    \begin{subfigure}{0.7\textwidth}
        \centering
        \includegraphics[width=0.6\textwidth]{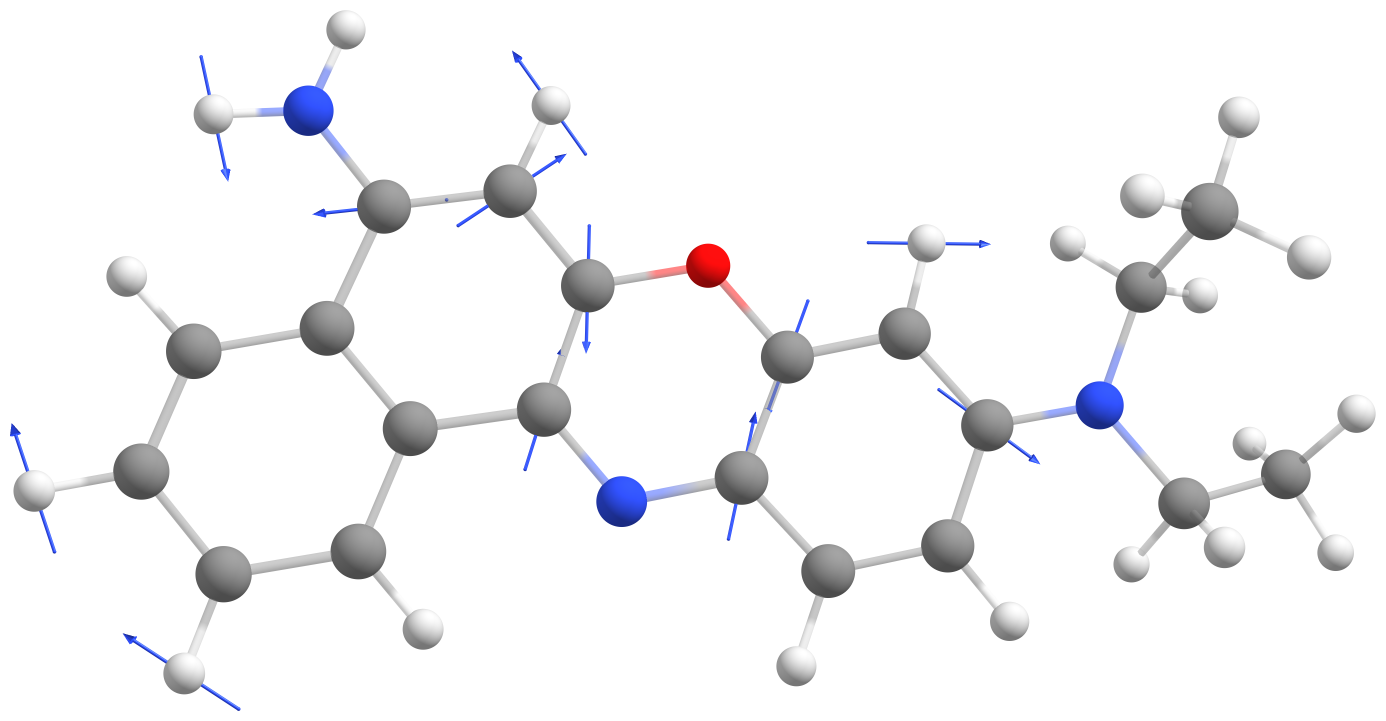}
        \caption{1517 cm$^{-1}$}
        \label{fig:nb_nm_1517}
    \end{subfigure}

    \begin{subfigure}{0.7\textwidth}
        \centering
        \includegraphics[width=0.6\textwidth]{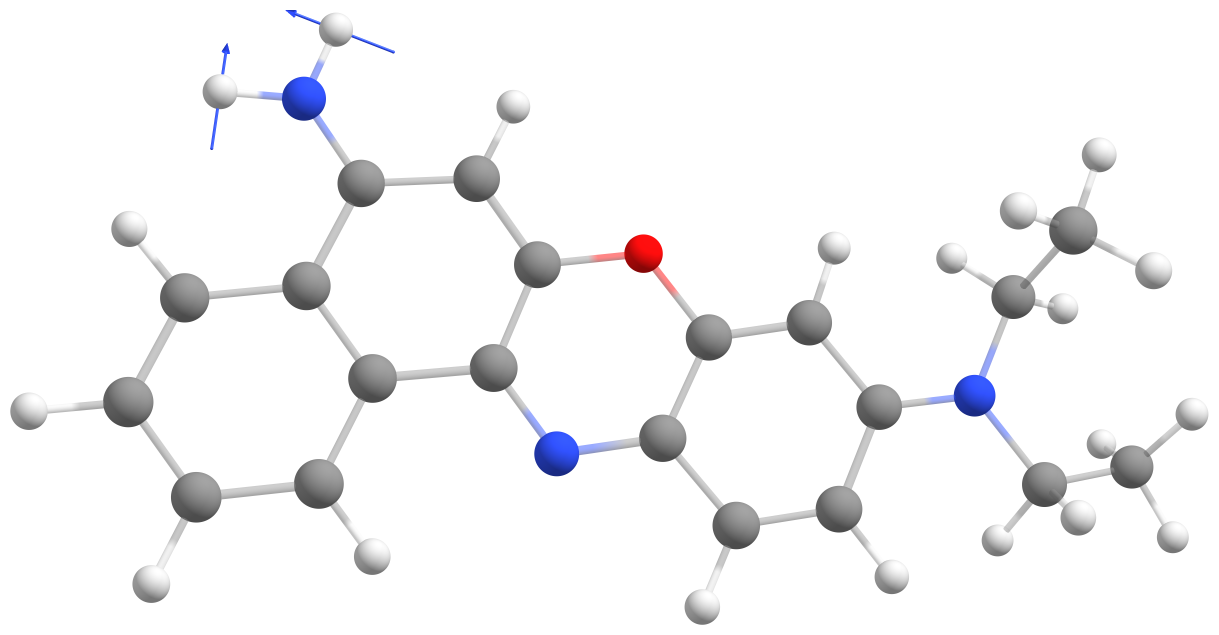}
        \caption{1636 cm$^{-1}$}
        \label{fig:nb_nm_1636}
    \end{subfigure}

    \caption{Ground state normal modes for the Nile blue chromophore in vacuum. Geometry optimization and frequency analysis were performed using TeraChem following the same DFT parameters used to obtain forces for the EqT machine learning potential outlined in SI Sec. \ref{sec:methods_grndEqT}.}
    \label{fig:nb_nm}
\end{figure}

SI Figure \ref{fig:nb_nm} shows three ground state normal modes for the Nile blue chromophore in vacuum at 583, 1517, and 1636 $\cm$. The first mode represents the sharp feature at 587 $\cm$ in the spectral density and is the ring-breathing mode of the central heteroatom-containing ring. The latter two modes include motion of the -NH$_2$ functional group involved in the hydrogen bonding analysis of SI Sec. \ref{sec:si_hbonds}. These normal modes are also used in the main text to elucidate how the interactions between the -NH$_2$ group and solvent molecules change the solvent-stripped spectral density. 

\newpage

\section{Comparison between the experimental and simulated pump-probe spectra}
\begin{figure}[h]
    \begin{center}\includegraphics[width=0.45\textwidth]{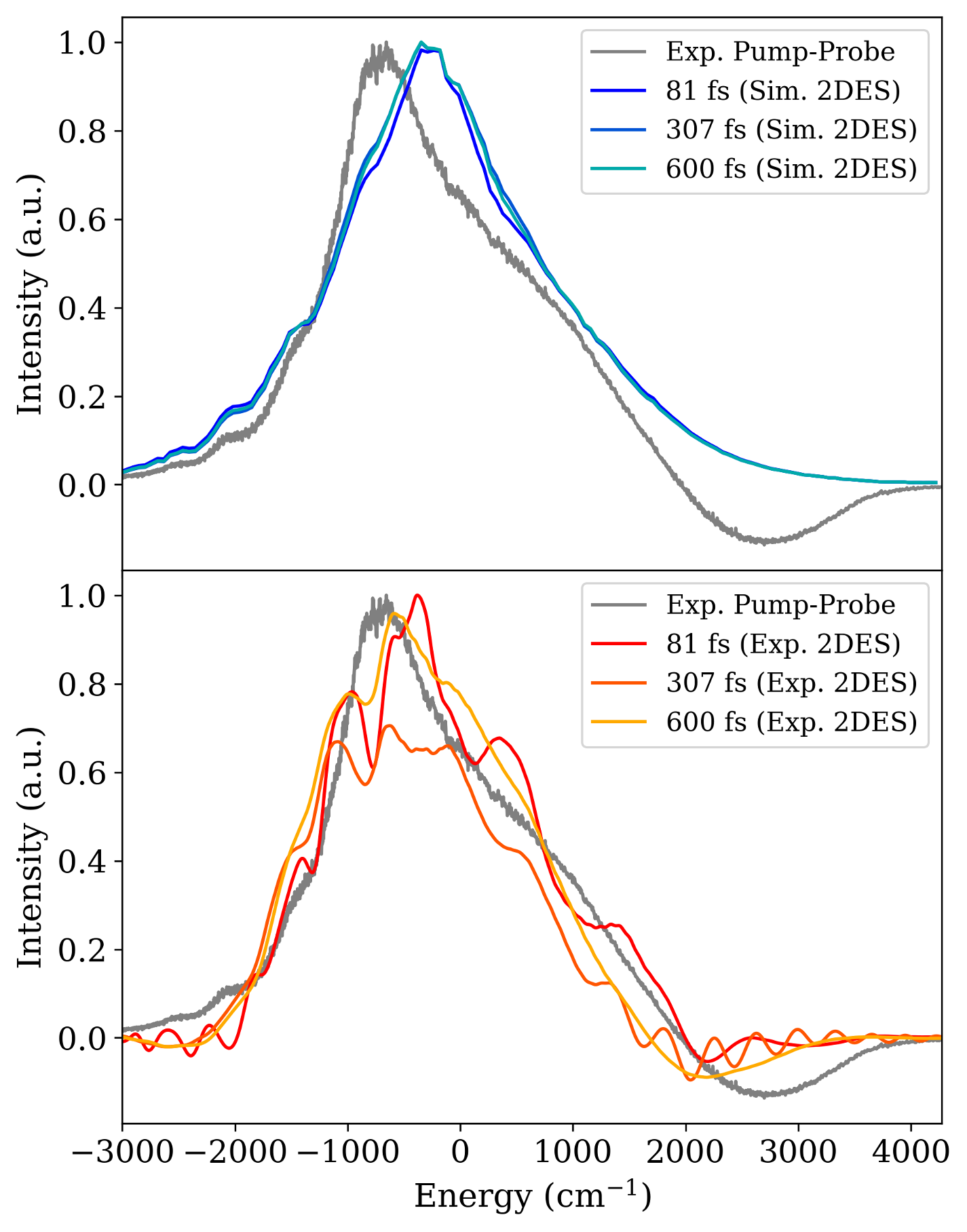}
    \end{center}
    \vspace{-5mm}
    \caption{Simulated pump-probe spectra obtain good agreement with experimental spectra. The experimental pump-probe spectrum obtained at a time delay of 3.34 ps (gray) is shown in both the top and bottom panels. The pump-probe spectra generated by integrating the simulated (top, blues) and experimental (bottom, reds) 2DES spectra at $t_2=81$, 307, and 600~fs are shown for comparison.} 
    \label{fig:pump_probe}
\end{figure}
The pump-probe spectra obtained from integrating the simulated (SI Fig. \ref{fig:pump_probe}, top) and experimental 2DES spectra (SI Fig. \ref{fig:pump_probe}, bottom) are in good agreement with each other. Both have similar line shapes and maxima just above the maximum of the directly measured pump-probe spectrum (gray).  The experimental pump-probe spectrum includes a strong negative feature above $\sim2000~\cm$ which is less prominent in the integrated experimental signal and absent from the simulation. While positive features in the pump-probe spectrum originate from stimulated emission and ground state bleach processes, negative features come from excited state absorption which is neglected in the simulations. 

\newpage
\section{Ground State Bleach and Stimulated Emission shifts without experimental pulse spectral profile}

\begin{figure}[h]
    \begin{center}
        \includegraphics[width=0.45\textwidth]{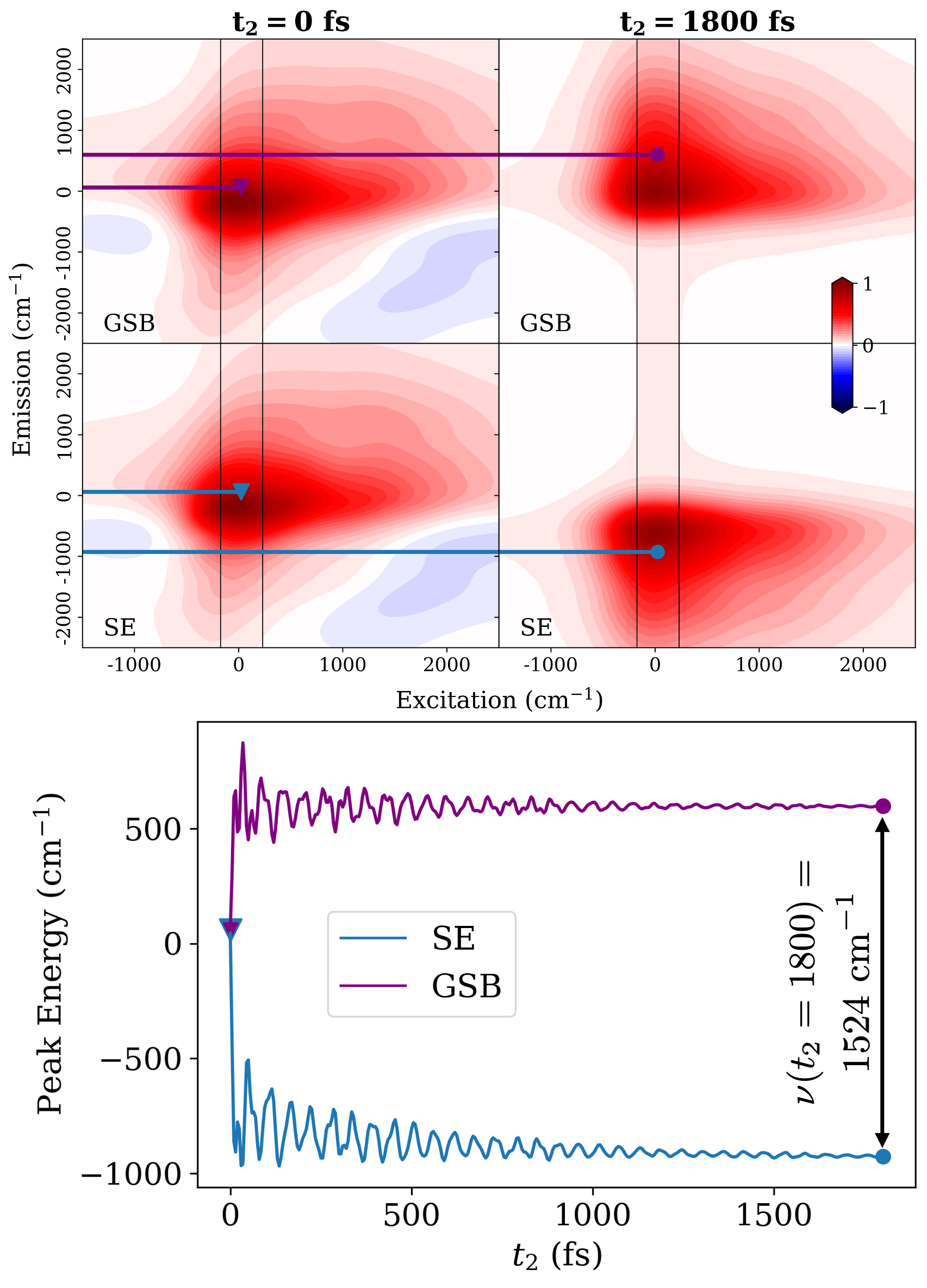}
    \end{center}
    \vspace{-5mm}
    \caption{Obtaining the Stokes shift and dynamic Stokes shift from our 2DES simulations without the experimental pulse spectral profile applied. Top panels: The simulated 2DES spectra with no experimental pulse spectral profile applied is decomposed into its SE and GSB components at two time delays (left column: $t_2=0$ fs and right column: $t_2=1800$ fs). These components are identical at $t_2=0$ fs and then separate as the delay time is increased due to the stabilization of the excited state. The weighted means (blue and purple markers) are obtained by averaging the data between the black dashed vertical lines and calculating the weighted mean emission frequency of the resulting distribution. Triangles and circles are used to show the position at $t_2=0~\mathrm{fs}$ and $t_2=1800~\mathrm{fs}$ respectively. Bottom panel: The position of the weighted means of the 2DES spectra for the SE and GSB  components as a function of the delay time. The separation of the two lines shown at each time is used to calculate the dynamic Stokes shift and the long time value (shown with double-headed arrow), \numericalresult{1524} $\cm$, is the Stokes shift.} 
    \label{fig:se_gsb_shifts}
\end{figure}
The Stokes shift analysis here is identical to that in Fig. 4 of the main text except here the experimental pulse spectral profile is not applied to the simulated spectra (SI Fig. \ref{fig:se_gsb_shifts}). As a result, the signals are broader and in particular, the SE component drops to lower frequencies when the pulse spectral profile is not applied. This leads to a long time splitting between the weighted average of the SE and GSB components of 1524 $\cm$ without the pulse spectral profile in contrast to the 1226 $\cm$ shift with the pulse spectral profile. The GSB signal shifts up in frequency largely due to vibronic coupling. At $t_2=0$, the GSB and SE components are identical but at longer delay times, the GSB component develops a vibronic progression to higher frequencies, analogous to a linear absorption spectrum while the SE component develops a vibronic progression to lower frequencies, analogous to a fluorescence spectrum.

\newpage
\section{Comparison of the Stokes shifts obtained from Experimental Absorption and Fluorescence Spectra}

\begin{figure}
    \centering
    \includegraphics[width=0.5\linewidth]{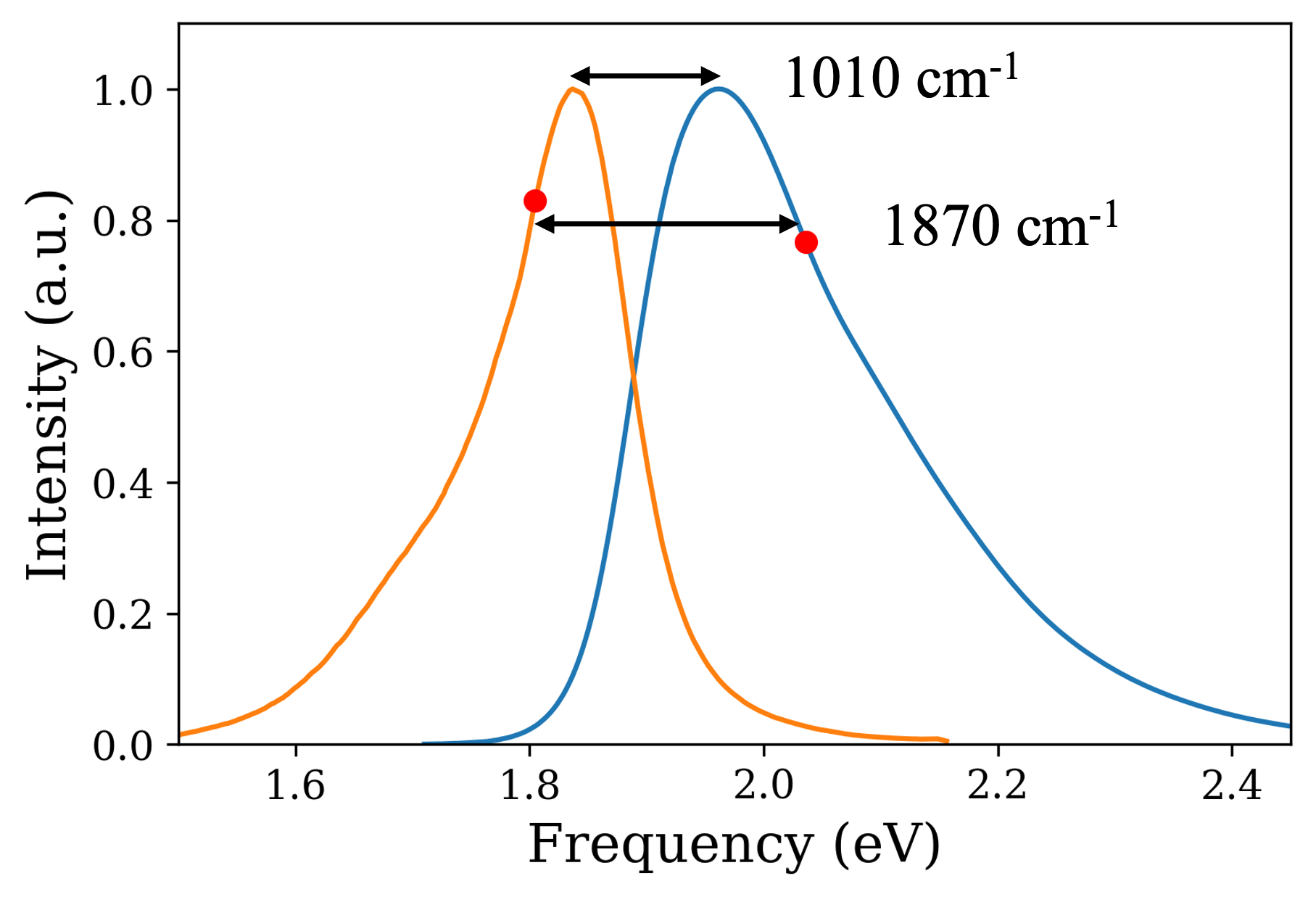}
    \caption{Experimental linear absorption and fluorescence spectra of Nile blue in ethanol. The absorption spectrum is in blue while fluorescence in orange. The weighted means of each spectra are marked with red dots.}
    \label{fig:linear_abs_fluor}
\end{figure}

The Stokes shift, the difference between the maxima of the absorption and fluorescence spectra, was measured to be 1010 $\cm$ using the experimental spectra of Nile blue in ethanol (SI Fig. \ref{fig:linear_abs_fluor}). However, the absorption and fluorescence spectra have tails that extend asymmetrically with the former extending to higher frequencies and the latter extending to lower frequencies. These asymmetries shift the weighted means to higher and lower frequencies respectively. By using the difference of the weighted means rather than maxima, the gap of 1870 $\cm$ reported in the main text is obtained. 

\newpage
\section{Reweighting to Obtain the Stokes Shift}
\label{si_sec:reweight}
In the main text, we provide a value for the difference between the weighted mean absorption and fluorescence spectra by reweighting the thermally sampled ground state configurations to account for their probabilities on the excited state. This section expands on how this value was obtained. The equilibrium ensemble average of an operator, $A$, for a system confined to a particular electronic state, $j$, is
\begin{align}
    \langle A \rangle_j = \frac{\int \mathrm{d}\mathbf{r}~ e^{-\beta U_j(\mathbf{r})}A(\mathbf{r}) }{\int \mathrm{d}\mathbf{r}~ e^{-\beta U_j(\mathbf{r})}}
\end{align}
where $\beta = 1/k_BT$ is the inverse temperature, $U_j(\mathbf{r})$ is the potential energy in ground ($j=g$) or excited ($j=e$) electronic state for a given molecular configuration and the integrals are over all position space. Note that the denominator is proportional to to the canonical partition function. If one defines the observable $A(\mathbf{r})$ to be the electronic energy gap, $\Delta U(\mathbf{r}) = U_e(\mathbf{r}) - U_g(\mathbf{r})$, then the ensemble average with respect to the $j^{th}$ electronic state of this observable is
\begin{align}
    \label{eq:weighted_average}
    \langle \Delta U \rangle_j = \frac{\int \mathrm{d}\mathbf{r}~e^{-\beta U_j(\mathbf{r})} (U_e(\mathbf{r}) - U_g(\mathbf{r})) }{\int \mathrm{d}\mathbf{r}~e^{-\beta U_j(\mathbf{r})}}.
\end{align}
 $\langle \Delta U \rangle_g$ is the thermally weighted average of the electronic energy gap obtained from a ground state simulation. To compute the change in the thermally weighted electronic energy gap between the ground and excited state ($\langle \Delta U \rangle_e-\langle \Delta U \rangle_g$), we also need the average energy gap evaluated on the excited electronic state, $\langle \Delta U \rangle_e$. Reweighting the ground state trajectory offers an approach to obtain $\langle \Delta U \rangle_e$ without performing dynamics on the excited state. By inserting the identities $1=e^{-\beta U_g(\mathbf{r})+\beta U_g(\mathbf{r})}$ and $1=\frac{\int \mathrm{d}\mathbf{r}~e^{-\beta U_g(\mathbf{r})}}{\int \mathrm{d}\mathbf{r}~e^{-\beta U_g(\mathbf{r})}}$, into Eq. \ref{eq:weighted_average}
\begin{align}
    \langle \Delta U \rangle_e &= \frac{\int \mathrm{d}\mathbf{r}~e^{-\beta U_e(\mathbf{r})} e^{-\beta U_g(\mathbf{r})+\beta U_g(\mathbf{r})}\Delta U(\mathbf{r}) }{\int \mathrm{d}\mathbf{r}~e^{-\beta U_e(\mathbf{r})}e^{-\beta U_g(\mathbf{r})+\beta U_g(\mathbf{r})}} \\
    &= \frac{\int \mathrm{d}\mathbf{r}~e^{-\beta U_g(\mathbf{r})} e^{-\beta \Delta U(\mathbf{r})}\Delta U(\mathbf{r}) }{\int \mathrm{d}\mathbf{r}~e^{-\beta U_g(\mathbf{r})}e^{-\beta \Delta U(\mathbf{r})}} \frac{\int \mathrm{d}\mathbf{r}~e^{-\beta U_g(\mathbf{r})}}{\int \mathrm{d}\mathbf{r}~e^{-\beta U_g(\mathbf{r})}} \\
    &= \frac{\langle e^{-\beta \Delta U}\Delta U \rangle_g}{\langle e^{-\beta \Delta U} \rangle_g}.
    \label{eq:reweighting}
\end{align}
Eq. \ref{eq:reweighting} thus provides an expression for the thermally averaged energy gap on the excited state in terms of properties sampled only on the ground state, assuming sufficient sampling on the ground state to converge the configurational average\cite{Zwanzig1954HighTemperatureGases, Tuckerman2010StatisticalSimulation, Ceriotti2012TheIntegration} . 

\newpage
\section{Analysis of hydrogen bonding sites on Nile blue}
\label{sec:si_hbonds}
\begin{figure}
    \centering
    \includegraphics[width=0.8\linewidth]{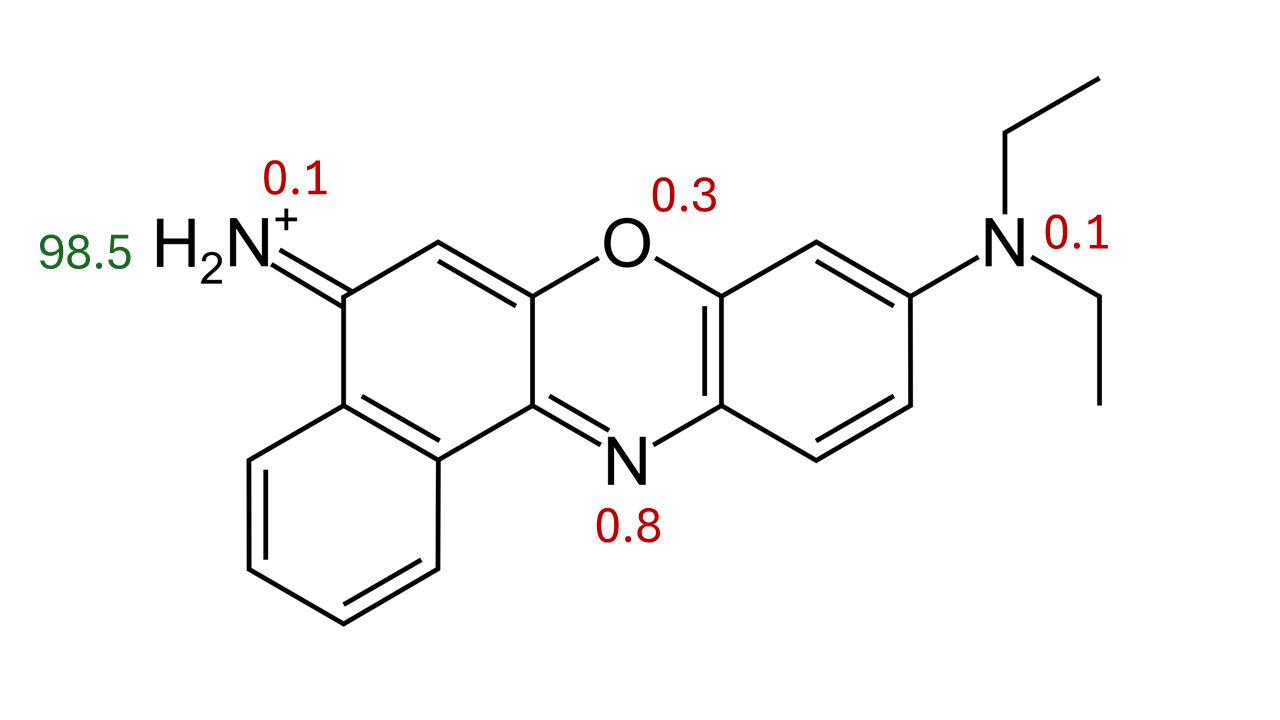}
    \caption{Percentage of configurations in total EqT trajectory with at least one hydrogen bonding solvent molecule at different chromophore atom sites. Primary amine group hydrogens on the Nile blue chromophore act as hydrogen bond donors to ethanol oxygen 98.5\% of the time during the trajectory, whereas all other chromophore sites act as hydrogen bond acceptors to ethanol hydroxyl hydrogen and experience minimal hydrogen bonding during the trajectory.}
    \label{fig:hbonds_EqT_traj}
\end{figure}

We define the presence of a hydrogen bond to occur when: i) the donor-acceptor heavy atom distance is less than 3.2 Å, and ii) the hydrogen-acceptor distance is less than 2.2 A, and iii) the donor–hydrogen–acceptor angle is greater than 110°.
The EqT dynamics of Nile blue in ethanol solvent present two types of hydrogen bonding interactions: those in which the heavy chromophore atoms act as hydrogen bond acceptors to the ethanol hydroxyl groups and those where the chromophore -NH$_2$ amine group donates a hydrogen bond to the ethanol oxygen. Figure \ref{fig:hbonds_EqT_traj} shows the site-specific breakdown of hydrogen bonding observed in configurations obtained from our 1 ns EqT trajectory. In this trajectory, 98.5\% of configurations show at least one instance of the Nile blue -NH$_2$ amine group donating a hydrogen bond to an adjacent ethanol solvent molecule, whereas all other forms of hydrogen bonding are only observed in a combined 1.3\% of configurations. 

\bibliography{references}